\documentclass[aps,prc,letterpaper,11pt,twoside,tightenlines,nofootinbib,showpacs,preprint]{revtex4}
\usepackage[latin1]{inputenc}
\usepackage{amsmath}
\usepackage{amsfonts}
\usepackage{amssymb}
\usepackage{graphicx}
\usepackage{rotating}

\begin{document}

\title{Effects of pseudoscalar-baryon channels in the dynamically generated vector-baryon resonances}

\author{E.~J.~Garzon$^1$ and E.~Oset$^1$}

\affiliation{$^1$ Departamento de F\'{\i}sica Te\'orica and IFIC, Centro Mixto Universidad de Valencia-CSIC,
Institutos de Investigaci\'on de Paterna, Aptdo. 22085, 46071 Valencia, Spain}

\begin{abstract} 
We study the interaction of vector mesons with the octet of stable baryons in the framework of the local hidden gauge formalism using a coupled channels unitary approach, including also the pseudoscalar-baryon channels which couple to the same quantum numbers. We examine the scattering amplitudes and their poles, which can be associated to known $J^P=1/2^-,3/2^-$ baryon resonances, and determine the role of the pseudoscalar-baryon channels, changing the width and eventually the mass of the resonances generated with only the basis of vector-baryon states.
\end{abstract}

\maketitle

\section{Introduction}
The approach to QCD at low energies using effective Lagrangians \cite{Weinberg:1978kz} has proved very valuable. In particular the implementation of chiral symmetry  \cite{Weinberg:1968de,Gasser:1983yg} has been a guiding principle to construct effective Lagrangians which account for the basic dynamics of the strong interaction and take the observable hadrons, mesons and baryons, as basic fields in those Lagrangians. Along this line, the use of chiral Lagrangians in combination with unitary techniques in coupled channels of mesons and baryons has been a very fruitful scheme to study the nature of many hadron resonances. It allows, among many other, to evaluate meson baryon scattering amplitudes, which sometimes show poles in the second Riemann sheet which are identified with existing baryon resonances. They are dynamically generated resonances. In this way the interaction of the octet of pseudoscalar mesons with the octet of stable baryons has lead to $J^P=1/2^-$ resonances which fit quite well the spectrum of the known low lying resonances with these quantum numbers \cite{Kaiser:1995cy,weise,Kaiser:1996js,angels,ollerulf,carmina,carmenjuan,hyodo,Hyodo:2006kg}.
Similarly, the interaction of the octet of pseudoscalar mesons with the decuplet of baryons also leads to many resonances that can be identified with existing ones of $J^P=3/2^-$ \cite{kolodecu,sarkar}. One interesting case is the one of the $\Lambda(1405)$, where all the chiral approaches find two poles close by, which have found experimental support in the analysis of Refs.~\cite{magas,sekihara}. 

Much work has been done using pseudoscalar mesons as building blocks, but the consideration of vectors instead of pseudoscalars is also receiving much attention lately. In the baryon sector the interaction of the $\rho$ $\Delta$ has been recently addressed in Ref.~\cite{vijande}, where three degenerate $N^*$ states around 1800 MeV and three degenerate $\Delta$ states around $1900$ MeV, with $J^P=1/2^-, 3/2^-, 5/2^-$, are found.
This work has been recently extended to the SU(3) space of vectors and baryons of the decuplet in Ref.~\cite{souravbao}.  The underlying theory for this study is the local hidden gauge formalism \cite{hidden1,hidden2,hidden3,hidden4}, which deals with the interaction of vector mesons and pseudoscalars in a way respecting chiral dynamics, providing the interaction of pseudoscalars among themselves, with vector mesons, and vector mesons among themselves. The theory provides the chiral Lagrangians as limiting cases at low energies through vector exchange diagrams. 

In the same line, the interaction of vector mesons with the octet of baryons has been addressed in Ref.~\cite{angelsvec}, where also many states are dynamically generated which can be associated to known resonances. A few other states remain as predictions or are difficult to associate to the known states which show a large dispersion from different experiments. More recently, work along this line has been done in Ref.~\cite{kanchan1}, and the concrete issue of the mixing of the pseudoscalar-baryon and vector-baryon channels, that we tackle here in the strangeness S=0 sector, has been addressed in Ref.~\cite{kanchan2} for S=-1. 
  
The next natural step in this direction is to put the pseudoscalar-baryon and vector-baryon states on the same footing and two works have been already done in this direction, the first one using a SU(6) scheme that invokes spin-isospin symmetry \cite{juansu6}. The transition from vector-baryon to pseudoscalar-baryon is implemented in this latter approach through the implicit exchange of an axial vector in the t-channel. Such a term is not present in the local hidden gauge approach, where instead, the exchange of pseudoscalar mesons coming from the vector-pseudoscalar-pseudoscalar vertex is responsible for the transition.
The second work \cite{kanchan2} investigates the mixing in the strangeness S=-1 sector, using a contact pseudoscalar-vector-baryon term obtained by gauging a theory with pseudoscalars and baryons.
In the present work we stick to the standard local hidden gauge formalism in the unitary gauge and study the mixing in a systematic way in order to see modifications to the resonances obtained in the work of Ref.~\cite{angelsvec} when the pseudoscalar-baryon channels are allowed to couple to the main building blocks of vector-baryon. 

The idea has already been used in Ref.~\cite{mishajuelich}, in the study of $\pi N$ scattering at intermediate energies, where the $\rho N$ channel is also included and a resonance is dynamically generated around 1700 MeV, which has the strongest coupling to the $\rho N$ channel.

The introduction of the pseudoscalar-baryon channels has as a main effect the widening of the resonances found in Ref.~\cite{angelsvec} and, except in some very particular case, has a negligible effect on the mass of the resonances. The consideration of the pseudoscalar-baryon channels also allows us to determine the partial decay width of the resonance into these channels and, by comparing with the PDG \cite{pdg} offers new elements to judge on the most appropriate association of the resonances found in Ref.~\cite{angelsvec} to the known resonances. 

\section{Formalism}
The part of the Lagrangian of the local hidden gauge approach that provides the interaction between vectors and is needed for the study of the vector-baryon interaction of Ref.~\cite{angelsvec}, is the three vector Lagrangian
\begin{equation}
{\cal L}^{(3V)}_{III}=ig\langle (\partial_\mu V_\nu -\partial_\nu V_\mu) V^\mu V^\nu\rangle
\label{l3V}\ ,
\end{equation}
where $V_\mu$ is the SU(3) matrix for the nonet of the $\rho$
\begin{equation}
V_\mu=\left(
\begin{array}{ccc}
\frac{\rho^0}{\sqrt{2}}+\frac{\omega}{\sqrt{2}}&\rho^+& K^{*+}\\
\rho^-& -\frac{\rho^0}{\sqrt{2}}+\frac{\omega}{\sqrt{2}}&K^{*0}\\
K^{*-}& \bar{K}^{*0}&\phi\\
\end{array}
\right)_\mu \ ,
\label{Vmu}
\end{equation}
and $g=\frac{M_V}{2 f}$, with $f$=93 MeV.
In the same way, the coupling of the vectors to pseudoscalar mesons is given by 
\begin{equation}
{\cal L}_{VPP}= -ig \langle [
P,\partial_{\nu}P]V^{\nu}\rangle \ ,
\label{lagrVpp}
\end{equation}
where here $P$ is the SU(3) matrix of the pseudoscalar mesons, 
\begin{equation}
P=\left(
\begin{array}{ccc}
\frac{\pi^0}{\sqrt{2}}+\frac{\eta_8}{\sqrt{6}}&\pi^+& K^{+}\\
\pi^-& -\frac{\pi^0}{\sqrt{2}}+\frac{\eta_8}{\sqrt{6}}&K^{0}\\
K^{-}& \bar{K}^{0}&-\frac{2}{\sqrt{6}} \eta_8\\
\end{array}
\right) \ .
\label{Pmatrix}
\end{equation}
As shown in Refs.~\cite{souravbao,angelsvec} the main source of vector-baryon interaction comes from the exchange of a vector meson in the t-channel between the vector and the baryon. This involves the ${\cal L}^{(3V)}_{III}$ Lagrangian of Eq.~(\ref{l3V}) and a Lagrangian for the coupling of the vector to the baryon, given by
\begin{equation}
{\cal L}_{BBV} =
g\left( \langle \bar{B}\gamma_{\mu}[V^{\mu},B]\rangle + 
\langle \bar{B}\gamma_{\mu}B \rangle \langle V^{\mu}\rangle \right) \ ,
\label{lbbv}
\end{equation}
where $B$ is now the SU(3) matrix of the baryon octet
\begin{equation}
B =
\left(
\begin{array}{ccc}
\frac{1}{\sqrt{2}} \Sigma^0 + \frac{1}{\sqrt{6}} \Lambda &
\Sigma^+ & p \\
\Sigma^- & - \frac{1}{\sqrt{2}} \Sigma^0 + \frac{1}{\sqrt{6}} \Lambda & n \\
\Xi^- & \Xi^0 & - \frac{2}{\sqrt{6}} \Lambda
\end{array}
\right) \ .
\end{equation}
For the transitions $V B \rightarrow P B$ we need the VPP Lagrangian of Eq.~(\ref{lagrVpp}) and one of the two pseudoscalar mesons is exchanged between the external VP and the baryon. The coupling of the pseudoscalar to the baryon for the two SU(3) octet is given by
\begin{equation}
{\cal L}_{BBP} =
\frac{F}{2} \langle \bar{B}\gamma_{\mu} \gamma_5 [u^{\mu},B]\rangle + 
\frac{D}{2} \langle \bar{B}\gamma_{\mu} \gamma_5 \left\lbrace u^{\mu}, B \right\rbrace \rangle
\label{lbbp}
\end{equation}
where $F=0.51$, $D=0.75$ \cite{Borasoy:1998pe} and at lowest order in the pseudoscalar field
\begin{equation}
u^{\mu}=-\frac{\sqrt{2}}{f} \partial^{\mu} P  \ ,
\label{umu}
\end{equation}
which allows to rewrite the Lagrangian of Eq.~(\ref{lbbp}) as
\begin{equation}
{\cal L}_{BBP} =
-\frac{\sqrt{2}}{f} \frac{D+F}{2} \langle \bar{B}\gamma_{\mu} \gamma_5 \partial^{\mu}P B \rangle  
-\frac{\sqrt{2}}{f} \frac{D-F}{2} \langle \bar{B}\gamma_{\mu} \gamma_5 B \partial^{\mu} P \rangle \ .
\label{lbbp2}
\end{equation}
Taking the SU(3) trace for a particular case of two baryons, the previous Lagrangian can be written in terms of an effective vertex as
\begin{equation}
-i t_{BBP} = \left\lbrace \alpha \frac{\left(D+F\right)}{2 f} + \beta \frac{\left(D-F\right)}{2 f}  \right\rbrace 
\vec{\sigma} \vec{k}
\label{lbbp3}
\end{equation}
where $\vec{k}$ is the incoming momentum of the meson in the BBP vertex. The coefficients $\alpha$ and $\beta$, can be found in Appendix \ref{appendix:BBPcoeff}.

\begin{figure}[ht!]
\begin{center}
\includegraphics[scale=0.7]{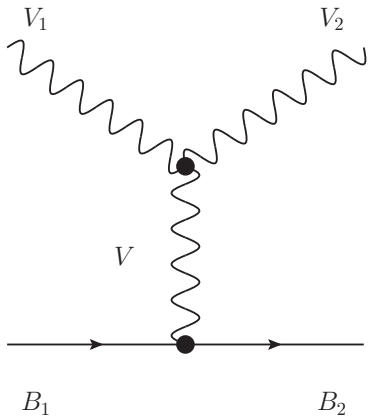}
\end{center}
\caption{Diagram of the $VB\rightarrow VB$ interaction mediated by a vector meson.}
\label{fig:vbvb}
\end{figure}
As shown in Ref.~\cite{angelsvec}, the leading term of the $V B \rightarrow V B$ interaction is given by the diagram of Fig.~\ref{fig:vbvb}
which involves the three vector vertex of Eq.~(\ref{l3V}), with one vector meson exchanged, and the coupling of this exchanged vector to the baryon, given by Eq.~(\ref{lbbv}). The potential provided by this term, keeping the dominant $\gamma ^0$ term in Eq.~(\ref{lbbv}), is given by
\begin{equation}
V_{i j}= - C_{i j} \, \frac{1}{4 f^2} \, \left( k^0 + k^\prime{}^0\right)~\vec{\epsilon}\,\vec{\epsilon }\,^\prime
\label{kernel}
\end{equation}
where $k^0, k^\prime{}^0$ are the energies of the incoming and outgoing vector mesons.
The result of Eq.~(\ref{kernel}) with the $\vec{\epsilon}\,\vec{\epsilon }\,^\prime$ factor for the polarization of the vector mesons stems from considering the three momentum of the external vectors small with respect to the mass of the vector mesons \cite{angelsvec}.
The $C_{i j}$ coefficients can be found in Appendix A of Ref.~\cite{angelsvec}, where the subindex $i$ and $j$ correspond to the different channels for all the states of isospin and strangeness.

However, it is more convenient to work with a relativistic potential as seen in Ref.~\cite{Oset:2001cn}, which is given by
\begin{equation}
V_{i j}= - C_{i j} \frac{1}{4 f^2} \left( 2 \sqrt{s} -M_{B_i} - M_{B_j} \right) 
			\left( \frac{M_{B_i}+E_{B_i}}{2 M_{B_i}} \right)^{1/2}
			\left( \frac{M_{B_j}+E_{B_j}}{2 M_{B_j}} \right)^{1/2}
\label{kernelrel}
\end{equation}
where $M_{B_i}$, $M_{B_j}$ are the masses of the initial or final baryons respectively, and $E_{B_i}$, $E_{B_j}$ their on shell energy.

This potential has been used as the input of the Bethe-Salpeter equation to study the scattering matrix, 
\begin{equation}
T = [1 - V \, G]^{-1}\, V
\label{eq:Bethe}
\end{equation}
where G is the loop function of a vector meson and a baryon which is calculated in dimensional regularization, as shown in Ref.~\cite{ollerulf,Oset:2001cn}, is given by
\begin{eqnarray}
G_{l} &=& i 2 M_l \int \frac{d^4 q}{(2 \pi)^4} \,
\frac{1}{(P-q)^2 - M_l^2 + i \epsilon} \, \frac{1}{q^2 - m^2_l + i
\epsilon}  \nonumber \\ &=& \frac{2 M_l}{16 \pi^2} \left\{ a_l(\mu) + \ln
\frac{M_l^2}{\mu^2} + \frac{m_l^2-M_l^2 + s}{2s} \ln \frac{m_l^2}{M_l^2} +
\right. \nonumber \\ 
& &  +
\frac{\bar{q}_l}{\sqrt{s}}
\left[
\ln(s-(M_l^2-m_l^2)+2\bar{q}_l\sqrt{s})+
\ln(s+(M_l^2-m_l^2)+2\bar{q}_l\sqrt{s}) \right. \nonumber  \\
& & \left. \left. - \ln(-s+(M_l^2-m_l^2)+2\bar{q}_l\sqrt{s})-
\ln(-s-(M_l^2-m_l^2)+2\bar{q}_l\sqrt{s}) \right]
\right\} \ ,
\label{eq:gpropdr}
\end{eqnarray}
where $\mu$=630 MeV is a regularization scale and 
$a_l(\mu)$ a subtraction constant with a value of -2 in Ref.~\cite{angelsvec}, which is considered a natural size in Ref.~\cite{ollerulf}.

In the cases where the iteration of the Bethe-Salpeter equation includes the $\rho$ or $K^*$ mesons, which have a relatively large widths, a convolution of the loop function $G$ with the mass distribution of these vector mesons is needed. So the loop function with the convolution would be
\begin{eqnarray}
\tilde{G}(s)= \frac{1}{N}\int^{(m+2\Gamma_i)^2}_{(m-2\Gamma_i)^2}d\tilde{m}^2
\left(-\frac{1}{\pi}\right) 
{\rm Im}\,\frac{1}{\tilde{m}^2-m^2+{\rm i} \tilde{m} \Gamma(\tilde{m})}
& G(s,\tilde{m}^2,\tilde{M}^2_B)\ ,
\label{Gconvolution}
\end{eqnarray}
where $\tilde{G}$ is normalized with
\begin{equation}
N=\int^{(m_\rho+2\Gamma_i)^2}_{(m_\rho-2\Gamma_i)^2}d\tilde{m}^2
\left(-\frac{1}{\pi}\right){\rm Im}\,\frac{1}{\tilde{m}^2-m^2_\rho+{\rm i} \tilde{m} \Gamma(\tilde{m})}
\label{Norm}
\end{equation}
considering the widths of the vectors $\Gamma_i$ ($i=\rho,~K^*$) $\Gamma_\rho$=149.4 MeV and $\Gamma_{K^*}$=50.5 MeV. The $\Gamma(\tilde{m})$ function is energy dependent and is given in Ref.~\cite{Geng:2008gx} as
\begin{equation}
\tilde{\Gamma}(\tilde{m})=\Gamma_i\frac{q^3_\mathrm{off}}{q^3_\mathrm{on}}\theta(\tilde{m}-m_1-m_2)
\end{equation}
with $m_1=m_2=m_\pi$ for the $\rho$ using that
\begin{equation}
q_\mathrm{off}=\frac{\lambda(\tilde{m}^2,m_\pi^2,m_\pi^2)}{2\tilde{m}},\quad
q_\mathrm{on}=\frac{\lambda(m_\rho^2,m_\pi^2,m_\pi^2)}{2 m_\rho}
\end{equation}
or $m_1=m_\pi$ and $m_2=m_K$ for the $K^*$ using
\begin{equation}
q_\mathrm{off}=\frac{\lambda(\tilde{m}^2,m_K^2,m_\pi^2)}{2\tilde{m}},\quad
q_\mathrm{on}=\frac{\lambda(m_{K^*}^2,m_K^2,m_\pi^2)}{2 m_{K^*}},
\end{equation}
where $\lambda$ is the K\"allen function and $\Gamma_i$ is the nominal width of the $\rho$ or the $K^*$.
Without these convolutions, the peaks that we find in the scattering matrix in the channels where the $\rho$ or the $K^*$ are involved, have a zero width or are very narrow. But when the loop function $G$ is replaced by the convolution function $\tilde{G}$ using the correspondent mass distribution, those peaks acquire a substantial width.

The $\vec{\epsilon}\,\vec{\epsilon }\,^\prime$ factor involving the polarization of the vector mesons factorizes in all the iterations of the potential implicit in the Bethe-Salpeter equations, as a consequence of which there will be a degeneracy in the spin, $1/2^-$, $3/2^-$, for the resonances found for each isospin and strangeness.

Once the scattering matrix is evaluated some peaks appear that can be associated to states. Next step is to find the poles associated to those peaks, in order to obtain the couplings of the different channels to those states. The method used is to search poles in the second Riemann sheet, changing the momentum $\vec{q}$ to $-\vec{q}$ in the analytical formula of the $G$ function when $Re(\sqrt{s})$ is over the threshold of the corresponding channel. Using this method one can find poles, as $(M_R+i\Gamma /2)$, where the real part correspond to the mass of the resonance and the imaginary part is half of the width of this state. However the convolution of the $G$ function eventually can make the pole disappear in channels with the $\rho$ or the $K^*$ mesons. In this case one can study the amplitude in the real axis using that near the peak the T matrix will be as
\begin{equation}
T_{i j} = \frac{g_i g_j}{\sqrt{s}-M_R+i \Gamma /2}
\label{eq:polematrix}
\end{equation}
where $M_R$ is the position of the maximum  and $\Gamma$ the width at half-maximum. The couplings $g_i$ and $g_j$ are related to the channels which couple to this resonance. Then one can take the diagonal channel where the coupling is largest and obtain
\begin{equation}
|g_i|^2 = \frac{\Gamma}{2} \sqrt{|T_{ii}|^2}
\end{equation}
where the coupling $g_i$ has an arbitrary phase. With one coupling determined, we can obtain the other ones from the $T_{ij}$ matrices using Eq. (\ref{eq:polematrix}), given by
\begin{equation}
g_j=g_i \frac{T_{i j}(\sqrt(s)=M_R)}{T_{i i}(\sqrt(s)=M_R)} \ .
\label{eq:polerealtion}
\end{equation}
This procedure has been used to calculate all the couplings of all the states of the vector-baryon interaction in Ref.~\cite{angelsvec}, so this method that we will use in order to calculate our results in the present work.

Using this formalism, nine resonances are found in Ref.~\cite{angelsvec}, which are associated to known states of the PDG \cite{pdg}, through the isospin and strangeness and the pole position. However the widths obtained with this approach are smaller than the experimental ones.
This result leads us to think that there should be some other mechanisms which contribute to the vector meson - baryon interaction potential. 
Since vector mesons and baryons couple to pseudoscalar mesons, there can be diagrams where the interaction is mediated by pseudoscalar mesons. The next section is devoted to study such mechanisms.

\section{The box diagram}
\begin{figure}[ht!]
\begin{center}
\includegraphics[scale=1]{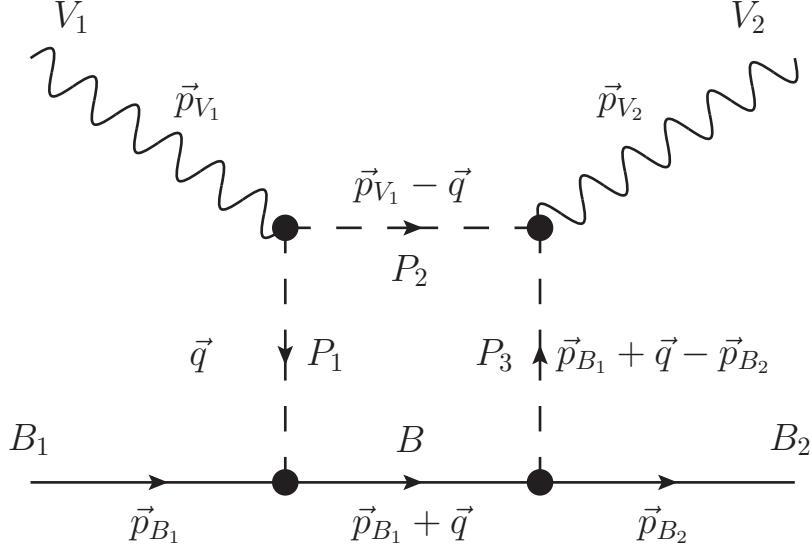}
\end{center}
\caption{Diagram of the $VB\rightarrow VB$ interaction mediated by a pseudoscalar meson-baryon loop.}
\label{fig:box}
\end{figure}
In addition to the driving term in the $VB\rightarrow VB$ potential of Eq.~(\ref{kernelrel}) there are other terms involving the exchange of pseudoscalar mesons that also contribute to this interaction. The idea is that an external vector meson decays into two pseudoscalar mesons, through the Lagrangian of Eq.~(\ref{lagrVpp}), and one of the pseudoscalar mesons is exchanged and absorbed by the baryon. Then a pseudoscalar-baryon state propagates in the intermediate state and the inverse procedure occurs in a second vertex, giving rise to a $VB$ again. The mechanism is depicted in Fig.~\ref{fig:box} in terms of a Feynman diagram, which gives a contribution to the $VB\rightarrow VB$ potential given by
\begin{eqnarray}
-i t_{Box}  &=& \int \frac{d^4 q}{(2 \pi)^4} i g C_{V_1} (\vec{p}_{V_1} -\vec{q} -\vec{q}) \cdot \vec{\epsilon_1} 
i g C_{V_2} (\vec{p}_{V_1} -\vec{q} - \vec{p}_{B_1} -\vec{q} + \vec{p}_{B_2}) \cdot \vec{\epsilon_2} \nonumber \\
	& & \frac{i}{q^2-m^2_1+i\epsilon} \frac{i}{(p_{V_1}-q)^2-m^2_2+i\epsilon} 
	\frac{i}{(p_{B_1}+q-p_{B_2})^2-m^2_3+i\epsilon} \nonumber \\
	& &	\frac{i}{(p_{B_1}+q)^0-E_B(\vec{p}_{B_1}+\vec{q})+i\epsilon}
	\vec{\sigma} \cdot \vec{q}~ \vec{\sigma} \cdot (-\vec{p_{B_1}}-\vec{q}+\vec{p_{B_2}} )
	 \nonumber \\
	& &  \left\lbrace \alpha_1 (D+F) +\beta_1 (D-F) \right\rbrace \frac{1}{2f} 
	\left\lbrace \alpha_2 (D+F) +\beta_2 (D-F) \right\rbrace \frac{1}{2f}	
\label{boxint0}
\end{eqnarray}
where $q$ is the loop four-momentum, $\vec{p}_{V_i}$ and $\vec{p}_{B_i}$ are the momenta of the external vector mesons and baryons and $m_i$ are the masses of the three pseudoscalar mesons of the loop. The coefficients of the VPP vertex $C_{V_i}$ are obtained from Eq.~(\ref{lagrVpp}) and the coefficients for the couplings of the pseudoscalars to the baryons, $\alpha_i$ and $\beta_i$ are shown in Table \ref{tab:BBPcoeff} of Appendix \ref{appendix:BBPcoeff}.

In other to calculate this integral, we perform analytically the integration over the $q^0$ component of the four-momentum of the loop. 
This leads to a residue, which is simplified to eliminate the fallacious poles with no determined position in the complex plane $((x-x_0+i\epsilon-i\epsilon^\prime)^{-1})$, and the result is given in Appendix \ref{appendix:boxint}.

Consistently with the approximation in Ref.~\cite{angelsvec} of neglecting the three momentum of the external vectors, we take here the same prescription and set these momenta to zero. The integral simplifies since we can substitute
\begin{equation}
\left\langle m' | \vec{\sigma}\cdot \vec{q} ~ \vec{\sigma}\cdot \vec{q} | m \right\rangle = 
\left\langle m' | \vec{q}\,^2 | m \right\rangle = \vec{q}\,^2 \delta_{m'm}
\end{equation}
and
\begin{equation}
q_i q_j \rightarrow \frac{1}{3}\vec{q}\,^2 \delta_{i j}
\end{equation}
The most generic expression for the vertex of Fig.~\ref{fig:box}, is given by
\begin{eqnarray}
V_{Box} &=&-g^2 C_{V_1} C_{V_2} \frac{1}{2 f^2}  
\left\lbrace a (D+F)^2+b (D-F)^2+ c (D+F)(D-F) \right\rbrace \vec{\epsilon_1} \vec{\epsilon_2} 
\nonumber \\
  & & \frac{1}{2 \pi^2} \frac{4}{3} \int d |\vec{q}| |\vec{q}|^6 \left( \frac{\Lambda^2}{\Lambda^2+|\vec{q}|^2}\right)^2 \frac{Num}{Den}
\label{boxint}
\end{eqnarray}
where we have introduced the usual form factor accompanying the Yukawa vertex with $\Lambda$=1 GeV.
The coefficients a, b and c, depend on each state, channel and particles involved in the box diagram, and can be found in Tables of Appendix \ref{appendix:boxintcoeff}. The numerator and denominator ($Num$ and $Dem$) of Eq.~(\ref{boxint}) are given in Appendix \ref{appendix:boxint}.

We found that for the channels with strangeness it was sufficient to consider only the diagonal terms of the Box diagram, but for the channels with S=0, the consideration of the non-diagonal terms made changes in the $K^* \Lambda \rightarrow K^* \Lambda$ amplitude that recommended its explicit consideration. Hence in the case of S=0, the non-diagonal terms are evaluated for the most relevant channels, $\rho N$, $K^* \Lambda$ and $K^* \Sigma$.

\begin{figure}[h]
\begin{center}
\includegraphics[scale=1]{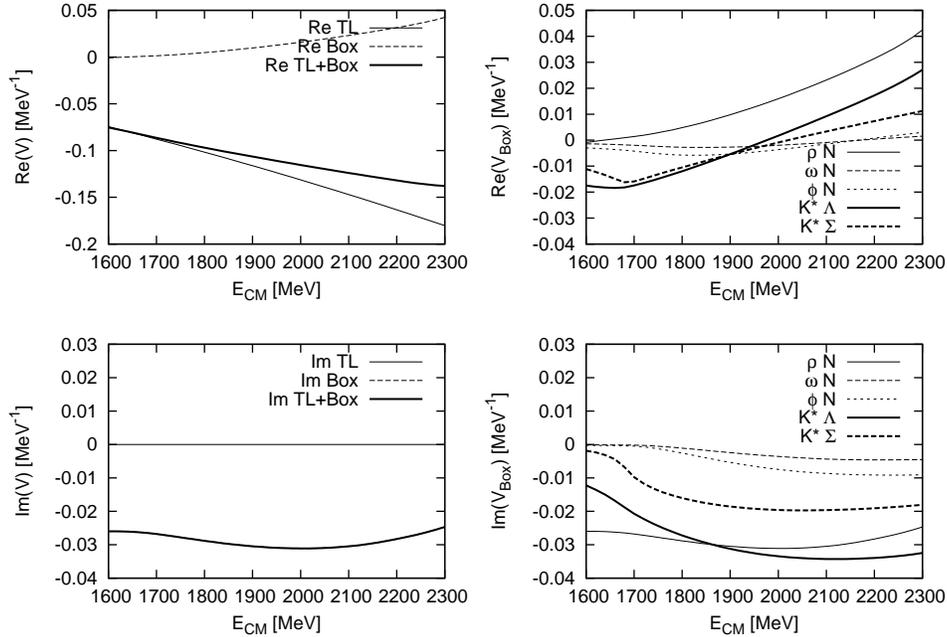}
\end{center}
\caption{Left) Comparison between the tree-level (TL) and the box diagram (Box) potential: (Top) Real part and (Bottom) Imaginary part. Right) Comparison of the box diagram potential between the different channels of the state I=1/2 and S=0: (Top) Real part and (Bottom) Imaginary part.}
\label{fig:comp}
\end{figure}

This $VB\rightarrow VB$ potential modifies the original potential, with only the vector meson exchange, giving a new potential including both interactions.
\begin{equation}
\tilde{V}=V+V_{Box}
\end{equation}
where V (tree level) and $V_{Box}$ are given by Eq.~(\ref{kernelrel}) and Eq.~(\ref{boxint}) respectively.
In Fig.~\ref{fig:comp} we can see a comparison of the real and imaginary parts of the tree level potential and the box potential for the $\rho N$ diagonal term. The same figure (right panel) shows the box contribution for the different diagonal terms of all channels that couple to $S=0,~I=1/2$. 

The results show that the box potential has a small real part, which is near zero close to the threshold, and an important imaginary part in contrast with the null imaginary part of the tree level potential. It is clear that the total potential will contribute to the scattering matrix generating a widening of the resonances.
Using this new potential, we can introduce it in the Bethe-Salpeter equation and recalculate the scattering matrix for all the states and channels.

Although there is a Lagrangian which involves the coupling of the pseudoscalar mesons connecting the baryon octet and  baryon decuplet, in the present work we have neglected the diagrams with a decuplet baryon as an intermediate state since the results of the box integral for these cases are less than 10$\%$ of those for the octet baryon in the intermediate state.

One can see qualitatively why this box diagram gives a small contribution to the real part. For this exercise we select the $\rho^+n\rightarrow\pi^+n$ vertex, together with the $\rho^+n\rightarrow\rho^+n$. The $\pi^0$ exchange diagram in the first case provides a transition amplitude 
\begin{equation}
t_{\rho^+n\rightarrow\pi^+n} \approx \frac{2 \sqrt{2}}{\sqrt{3}} \frac{\vec{q}~^2}{{q^0}^2-\vec{q}~^2-m_\pi^2} g \frac{D+F}{2f}
\end{equation}
while the $\rho^+n\rightarrow\rho^+n$ gives
\begin{equation}
t_{\rho^+n\rightarrow\rho^+n} \approx \frac{2 m_\rho}{4f^2}
\end{equation}
and we chose $q^0$, $\vec{q}$ in the $\pi$ exchange corresponding to the $\pi N$ on shell situation for $\sqrt{s} \approx 1650$ MeV. We find the $\rho^+n\rightarrow\rho^+n$ vertex about 4 times bigger than the $\rho^+n\rightarrow\pi^+n$ one. Yet, one must consider that to get a contribution to $\rho^+n\rightarrow\rho^+n$, the $\rho^+n\rightarrow\pi^+n$ terms must be iterated in the box (and multiplied by $G(\pi,n)$) and this further reduces the contribution of the box diagram to the real part of the $VN\rightarrow VN$ amplitude. The accurate numerical results can be seen in Fig.~\ref{fig:comp}, corroborating the qualitative explanation.

\section{Contact terms VPBB for the box diagrams}
\begin{figure}[ht!]
\begin{center}
\includegraphics[scale=0.5]{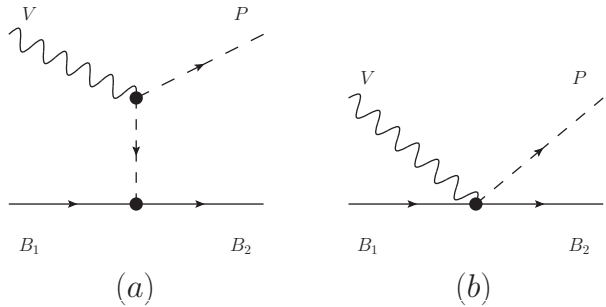}
\end{center}
\caption{Diagram of the $VB\rightarrow PB$ contact term.}
\label{fig:vbpb}
\end{figure}

As we are following the local hidden gauge, there are some other diagrams requested by the gauge invariance of the theory in the presence of baryons. Hence, in addition to the diagram of Fig.\ref{fig:vbpb} (a), we have to add the diagram of Fig.\ref{fig:vbpb} (b). This has been known for long and has been taken into account in the literature under the denomination of vertex corrections \cite{Klingl:1997kf,Chanfray:1993ue,Herrmann:1993za,Antinori:2000ph,Cabrera:2000dx,Oset:2000eg,kanchan2}. This diagram of Fig.\ref{fig:vbpb} (b) corresponds to a contact term of VPBB and as seen in Ref.~\cite{Cabrera:2000dx}, the correspondence between the meson in flight term (Fig.\ref{fig:vbpb} (a)) and the contact term (Fig.\ref{fig:vbpb} (b)) is given by:

Meson in flight:
\begin{equation}
-i t_{con}=C_V g \vec{\epsilon} (\vec{P}_{V_1}+\vec{q}+\vec{q}) \frac{1}{(P_{V_1}+q)^2-m^2} \left\lbrace  \alpha \left( D+ F \right) + \beta \left( D-F \right) \right\rbrace \frac{1}{2 f}
\vec{\sigma} (\vec{P}_{V_1}+\vec{q})
\end{equation}

Contact term:
\begin{equation}
-i t_{con}=C_V g \left\lbrace  \alpha \left( D+ F \right) + \beta \left( D-F \right) \right\rbrace \frac{1}{2 f}
\vec{\sigma} \vec{\epsilon}
\label{eq:contactvertex}
\end{equation}

\begin{figure}[h]
\begin{center}
\includegraphics[scale=0.5]{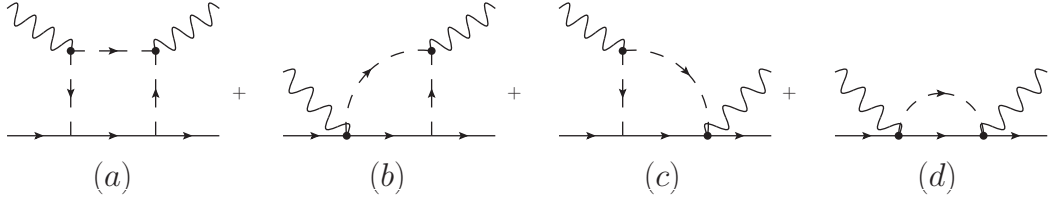}
\end{center}
\caption{Diagrams of the vector-baryon interaction mediated by pseudoscalars channels.}
\label{fig:diagramscontact}
\end{figure}

Using this vertex, one has to rewrite Eq.~(\ref{boxint0}), to calculate the diagrams of Fig.~\ref{fig:diagramscontact}. First of all, we should relabel the momenta of the the box diagram in a more convenient way, 

\begin{eqnarray}
-i t_{Box} &=& \int \frac{d^4 q}{(2 \pi)^4} 
		i g C_{V_1} (-\vec{P}_{V_1} -\vec{q}-\vec{q} ) \vec{\epsilon}_1
		i g C_{V_2} (-\vec{q} -\vec{P}_{V_2} -\vec{q} ) \vec{\epsilon}_2
		\nonumber \\
		& & \frac{i}{(P_{V_1}+q)^2-m_1^2} \frac{i}{q^2-m^2} \frac{i}{(P_{V_1}-q)^2-m_3^2}
		\nonumber \\
		& & \frac{i}{(P_{B_1}^0+P_{V_1}^0+q^0)-E_{B_1}(\vec{q})+i\epsilon} \frac{M_{B_1}}{E_{B_1}(q)}
		\vec{\sigma} (\vec{P}_{V_1} +\vec{q} ) \vec{\sigma} (-\vec{P}_{V_2} -\vec{q} )
		\nonumber \\
		& & \left\lbrace \alpha_1 (D+F) +\beta_1 (D-F) \right\rbrace  
  			\left\lbrace \alpha_2 (D+F) +\beta_2 (D-F) \right\rbrace \frac{1}{4f^2}
\label{boxmod}
\end{eqnarray}

While the (a) diagram of Fig.~\ref{fig:diagramscontact} gives an equal contribution to spin 1/2 and 3/2, the diagrams of Figs.~\ref{fig:diagramscontact} (b), (c) and (d) only contribute to spin 1/2. This is because the operator $\vec{\sigma} \vec{\epsilon}$ only couples to spin 1/2. This is done explicitly in Appendix \ref{appendix:spin}, where we show that the matrix element of $\vec{\sigma} \vec{\epsilon}$ is $\sqrt{3}~\delta_{J, 1/2}$. Intuitively we can see that only J=1/2 is allowed by looking at the intermediate PB state, necessarily in s-wave because of the $\vec{\sigma} \vec{\epsilon}$ coupling, hence the total spin J is the spin carried by the nucleon, J=1/2.

Using the vertex of Eq.~(\ref{eq:contactvertex}) in the expression of the Box given by Eq.~(\ref{boxmod}), we can evaluate the diagram $b$ of Fig.~\ref{fig:diagramscontact}, and obtain

\begin{eqnarray}
-i t^{CP}_{Box}  &=& g C_{V_1} g C_{V_2} 
  \left\lbrace \alpha_1 (D+F) +\beta_1 (D-F) \right\rbrace  
  \left\lbrace \alpha_2 (D+F) +\beta_2 (D-F) \right\rbrace \frac{1}{4f^2} \nonumber \\
 	& & 
 	\int \frac{d^3 q}{(2 \pi)^3} \vec{q}~^2 \frac{M_{B_1}}{E_{B_1}(q)}
 	\frac{1}{\omega_2 (q)} \frac{1}{\omega_3 (q)}
 	 \frac{1}{P_{V_2}^0+\omega_2 (q)+\omega_3 (q)}
 	\nonumber \\ 	
 	& &
 	 \frac{1}{P_{V_2}^0-\omega_2 (q)-\omega_3 (q)+i\epsilon}
 	  	\frac{1}{P_{B_1}^0 + P_{V_1}^0-\omega_2 (q)-E_{B_1} (q) +i\epsilon}
\nonumber \\ 	
 	& &
  	\frac{1}{P_{B_1}^0 + P_{V_1}^0-P_{V_2}^0 (q)-\omega_3 (q)-E_{B_1} (q) +i\epsilon}
  	 \left( \frac{\Lambda ^2}{\Lambda ^2+ \vec{q} ^2} \right)^2		
\nonumber \\ 	
 	& &
 	\left\lbrace (\omega_2(q)+\omega_3(q))(P_{B_1}^0-\omega_2(q)-E_{B_1}(q)-\omega_3(q)+P_{V_1}^0)
 	-\omega_3(q)P_{V_2}^0 \right\rbrace	
\label{boxcp}
\end{eqnarray}
In the same way, for the evaluation of the diagram $c$ in Fig.~(\ref{fig:diagramscontact}), we substitute the term of the outgoing vector by the contact term. Hence the result is the same as in Eq.~\ref{boxcp} changing $\omega_3 \rightarrow \omega_1$ and $P_{V_2}^0 \rightarrow P_{V_1}^0$, and one gets
\begin{eqnarray}
-i t^{PC}_{Box}  &=& g C_{V_1} g C_{V_2} 
  \left\lbrace \alpha_1 (D+F) +\beta_1 (D-F) \right\rbrace  
  \left\lbrace \alpha_2 (D+F) +\beta_2 (D-F) \right\rbrace \frac{1}{4f^2} \nonumber \\
 	& & 
 	\int \frac{d^3 q}{(2 \pi)^3} \vec{q}~^2 \frac{M_{B_1}}{E_{B_1}(q)}
 	\frac{1}{\omega_2 (q)} \frac{1}{\omega_1 (q)}
 	 \frac{1}{P_{V_1}^0+\omega_2 (q)+\omega_1 (q)}
 	\nonumber \\ 	
 	& &
 	 \frac{1}{P_{V_1}^0-\omega_2 (q)-\omega_1 (q)+i\epsilon}
 	  	\frac{1}{P_{B_1}^0 + P_{V_2}^0-\omega_2 (q)-E_{B_1} (q) +i\epsilon}
\nonumber \\ 	
 	& &
  	\frac{1}{P_{B_1}^0 + P_{V_2}^0-P_{V_1}^0 -\omega_1 (q)-E_{B_1} (q) +i\epsilon}
  	 \left( \frac{\Lambda ^2}{\Lambda ^2+ \vec{q} ^2} \right)^2		
\nonumber \\ 	
 	& &
 	\left\lbrace (\omega_2(q)+\omega_1(q))(P_{B_1}^0-\omega_2(q)-E_{B_1}(q)-\omega_1(q)+P_{V_2}^0)
 	-\omega_1(q)P_{V_1}^0 \right\rbrace		
\label{boxpc}
\end{eqnarray}

Finally, the last diagram has two contact terms, as shown in Fig.\ref{fig:diagramscontact} (d), and the expression of the t-matrix in this case is given by
\begin{eqnarray}
-i t^{CC}_{Box}  &=& g C_{V_1} g C_{V_2} 
  \left\lbrace \alpha_1 (D+F) +\beta_1 (D-F) \right\rbrace  
  \left\lbrace \alpha_2 (D+F) +\beta_2 (D-F) \right\rbrace \frac{1}{4f^2} \nonumber \\
 	& & 
 	3 \int \frac{d^3 q}{(2 \pi)^3} \frac{M_{B_1}}{E_{B_1}(q)}
 	\frac{1}{2 \omega_2 (q)} 
 %	\nonumber \\ 	
% 	& &
  	\frac{1}{P_{B_1}^0 + P_{V_1}^0-\omega_2 (q)-E_{B_1} (q) +i\epsilon}
  	 \left( \frac{\Lambda ^2}{\Lambda ^2+ \vec{q} ^2} \right)^2		
% \nonumber \\ 	
\label{boxcc}
\end{eqnarray}

These diagrams should be included in the potential in order to evaluate the scattering matrix.

\begin{equation}
\tilde{V}=V+V_{Box}+(V_{Box}^{CP}+V_{Box}^{PC}+V_{Box}^{CC})\delta_{J,1/2}
\label{vertexcontact}
\end{equation}

The presence of the terms $V_{Box}^{CP}$, $V_{Box}^{PC}$ and $V_{Box}^{CC}$ breaks the degeneracy in total spin J, that one has with the terms $V$ and $V_{Box}$.

One finds that the contact term, or Kroll-Ruderman in the nomenclature of Ref.~\cite{kanchan2}, is more important than the pseudoscalar exchange, around 1.5-2.5 times bigger depending on the cases, and has opposite sign. In Ref.~\cite{kanchan2} only this term was used but not the pseudoscalar exchange term. As a consequence, the corrections due to the VB and PB mixing only affect the J=1/2 case in the work of Ref.~\cite{kanchan2}.
As we find here, the corrections from the mixing due to the pseudoscalar exchange term in J=3/2 are relatively small, justifying its neglect in Ref.~\cite{kanchan2}. On the other hand for J=1/2, only the contact term is used in Ref.~\cite{kanchan2} and the destructive interference with the pseudoscalar exchange term is missed. This could justify why the effects due to the mixing in the scattering amplitudes of PB found in Ref.~\cite{kanchan2} are bigger than those found here for the VB case.

There are other possible couplings in the approach, like the consideration of the tensor coupling for the VBB vertex used in Ref.~\cite{kanchan1}, which was found to have minor effects in the present context in Ref.~\cite{kstmed}. Other couplings concerning vectors have also been exploited, like using the anomalous coupling of VVP. This was done in Ref.~\cite{vijande} and found to provide negligible corrections in the problem of $\Delta \rho$ interaction. One exception where this term turned out to be relatively important was in the study of the $\eta^\prime N$ interaction and its mixing with the VB channels. Indeed, the consideration of the anomalous $\eta^\prime K^* \bar{K}^*$ in Ref.~\cite{etaangels} was found relatively relevant, but only because the $\eta^\prime PV$ normal coupling of Eq.~(\ref{lagrVpp}) is zero when the $\eta^\prime$ is considered as a SU(3) singlet in a first approximation. 

\section{Results}
\begin{figure}[ht!]
\begin{center}
\includegraphics[scale=1.5]{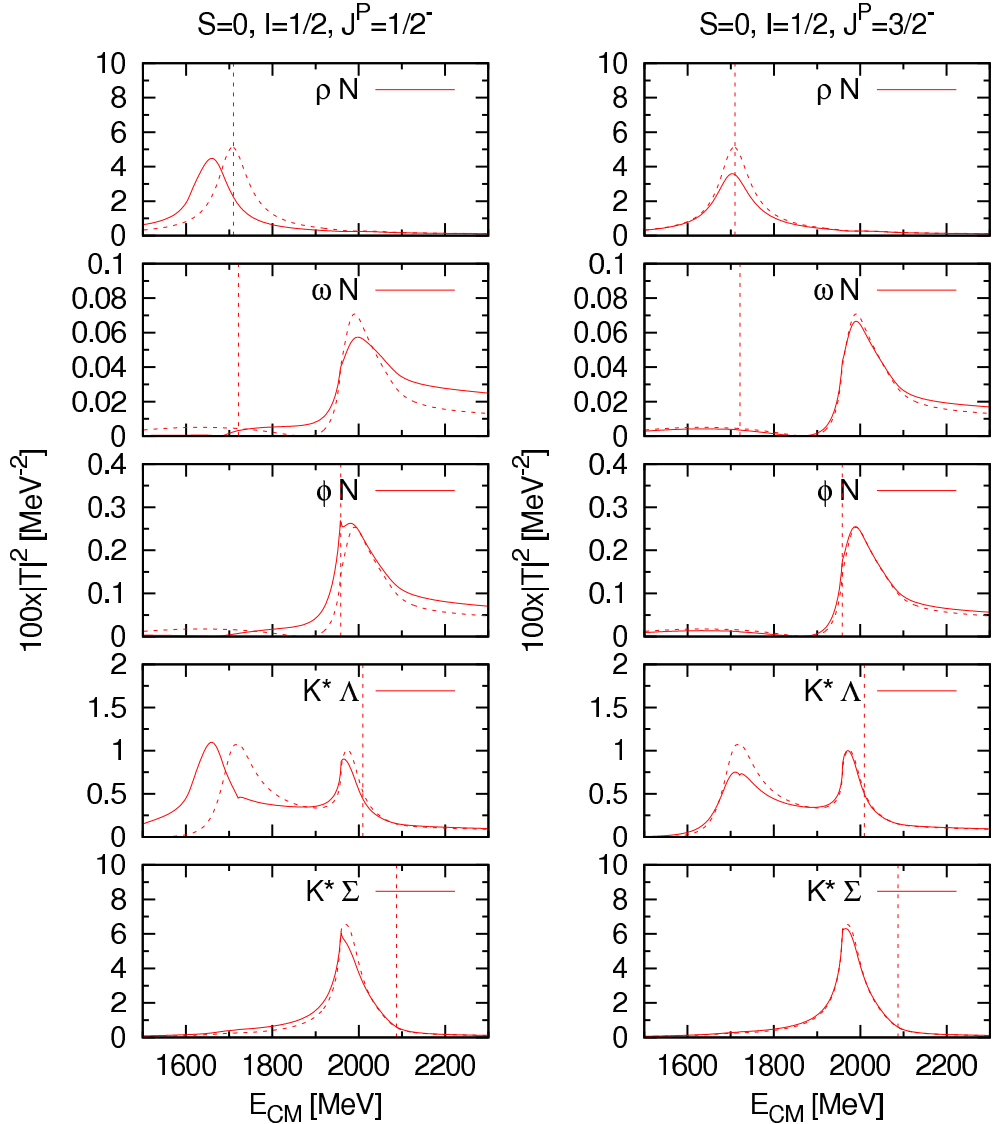} 
\end{center}
\caption{$|T|^2$ for the S=0, I=1/2 states. Dashed lines correspond to tree level only and solid lines are calculated including the box diagram potential. Vertical dashed lines indicate the channel threshold.}
\label{res1}
\end{figure}

\begin{figure}[ht!]
\begin{center}
\includegraphics[scale=1.5]{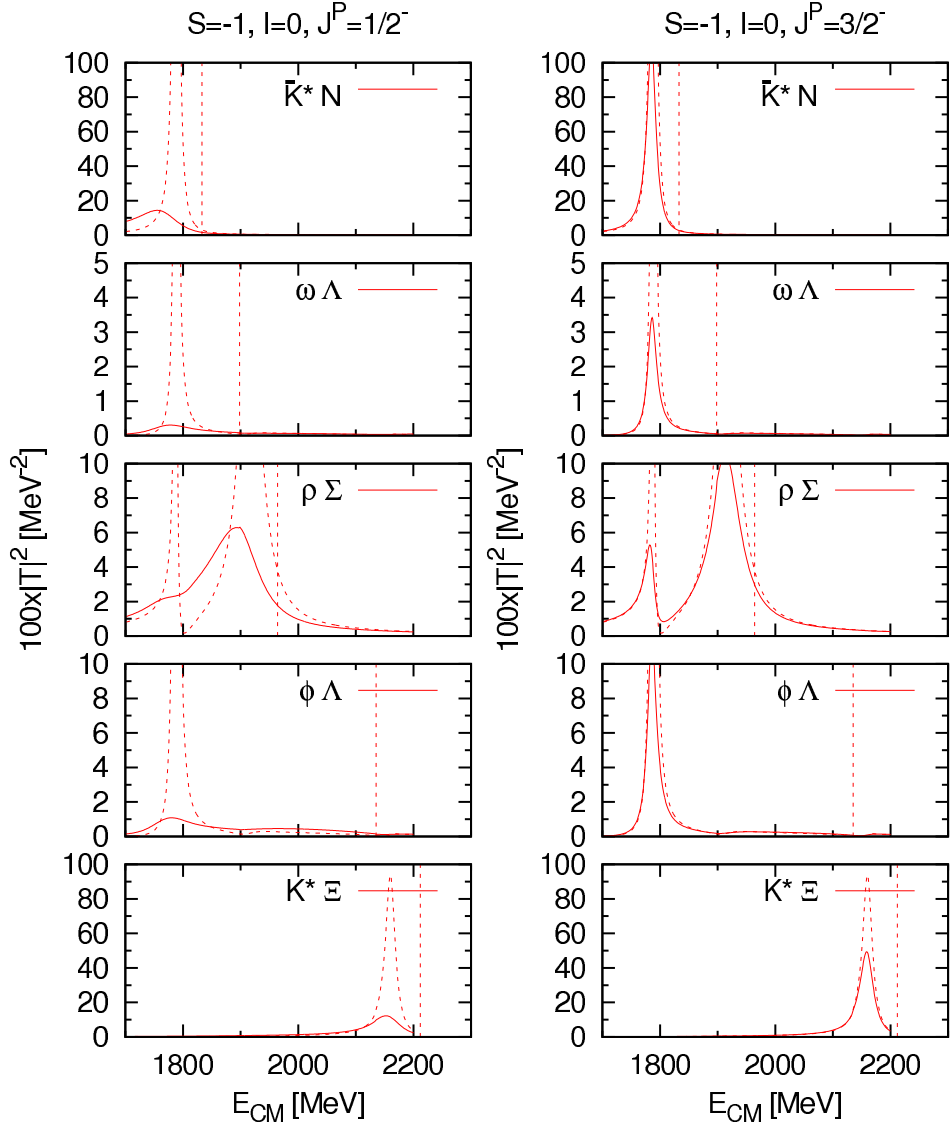} 
\end{center}
\caption{$|T|^2$ for the S=-1, I=0 states. Dashed lines correspond to tree level only and solid lines are calculated including the box diagram potential. Vertical dashed lines indicate the channel threshold.}
\label{res2}
\end{figure}

\begin{figure}[ht!]
\begin{center}
\includegraphics[scale=1.5]{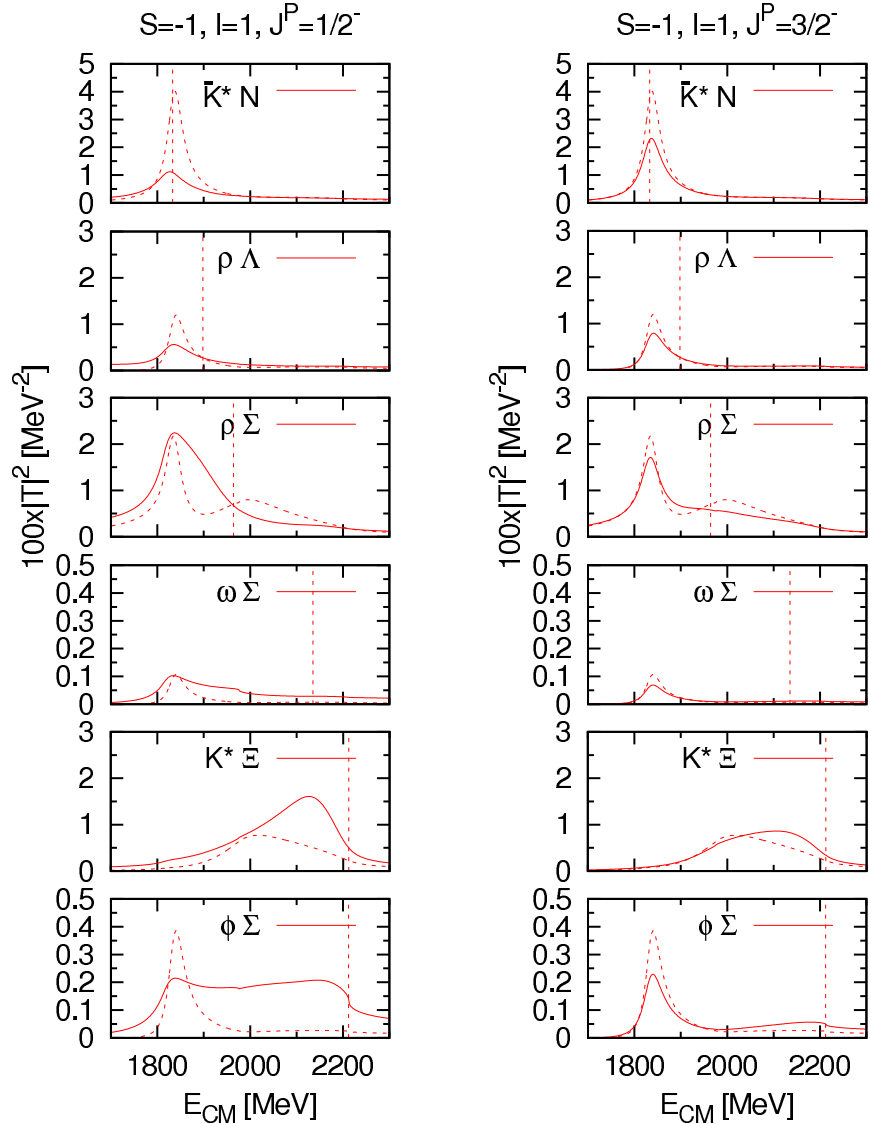} 
\end{center}
\caption{$|T|^2$ for the S=-1, I=1 states. Dashed lines correspond to tree level only and solid lines are calculated including the box diagram potential. Vertical dashed lines indicate the channel threshold.}
\label{res3}
\end{figure}

\begin{figure}[ht!]
\begin{center}
\includegraphics[scale=1.5]{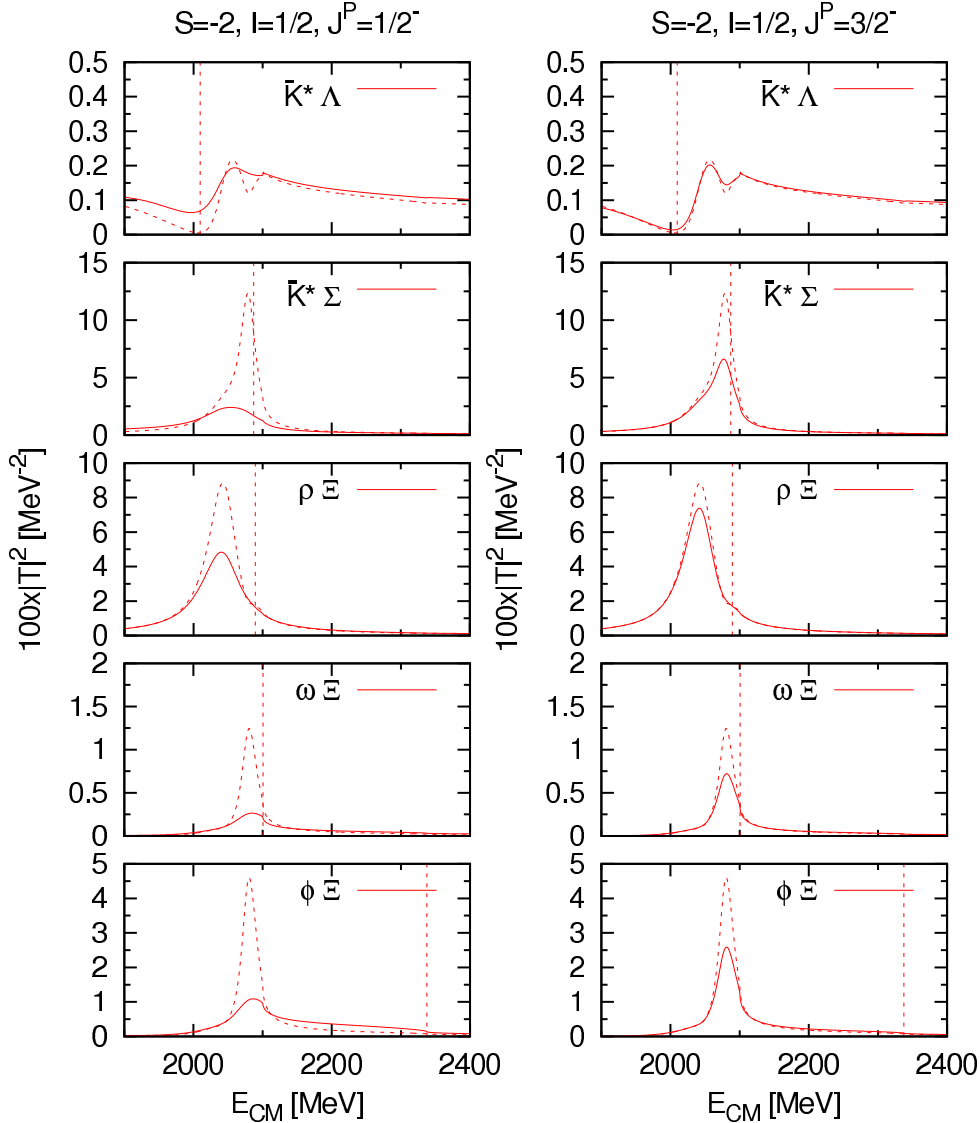} 
\end{center}
\caption{$|T|^2$ for the S=-2, I=1/2 states. Dashed lines correspond to tree level only and solid lines are calculated including the box diagram potential. Vertical dashed lines indicate the channel threshold.}
\label{res4}
\end{figure}

The use of the Bethe-Salpeter equation, generates the following scattering matrices for the different states, shown in Figs. \ref{res1}, \ref{res2}, \ref{res3} and \ref{res4}. We present the results of $|T_{ii}|^2$ as a function of $\sqrt{s}$.
We consider interesting the comparison of the scattering matrix obtained using only the potential of the vector interaction, and the new results of the Bethe-Salpeter equation with the combined potential. In all the figures, we present both results, the first one as a dashed line and the last one with a solid line.
The most striking feature in all the figures when including the intermediate PB states is, as expected, an increase in the width of the resonances.
However, in some cases we also observe a shift of the peak for J=1/2 of some resonances. This is the case for the resonances seen for $J^P=1/2^-$ in the $\rho N$ and $K^* \Lambda$ channels in Fig.~\ref{res1}.

%\subsection{I=1/2, S=0}
In Fig.~\ref{res1} we see two peaks for the state of $S=0$ and $I=1/2$, one around 1700 MeV, in channels $\rho N$ and $K^* \Lambda$, and another peak near 1980 MeV, which appears in all the channels except for $\rho N$. 
We can see that the mixing of the PB channels affects differently the two spins, $J^P=1/2^-$ and $3/2^-$, as a consequence of the extra mechanisms contributing to the $J^P=1/2^-$ case discussed in the previous section. The effect of the box diagram on the $J^P=3/2^-$ sector is small, however the PB-VB mixing mechanism are more important in the $J^P=1/2^-$ sector. The most important feature is a shift of the peak around 1700 MeV, which appears now around 1650 MeV. This breaking of the degeneracy is most welcome since this allows us to associate the $1/2^-$ peak found at 1650 MeV with the $N^*(1650)(1/2^-)$ while the peak for $3/2^-$ at 1700 MeV can be naturally associated to the $N^*(1700)(3/2^-)$. We shall discuss the other peak in the next section.

%\subsection{I=0, S=-1}
Fig.~\ref{res2} shows the results for $S=-1$, with $I=0$. The left column corresponds to $J^P=1/2^-$ and the right column to $J^P=3/2^-$. We can find three peaks for these quantum numbers. The first one around 1780 MeV  appears in the channels $\bar{K}^*N$, $\omega \Lambda$, $\rho \Sigma$ and $\phi \Lambda$, but not for the channel $K^* \Xi$. The second peak appears only in the $\rho \Sigma$ channel around 1900 MeV. The third peak is near 2150 MeV, an is only visible in the $K^* \Xi$ channel. Once again we see the different effects of the PB-VB mixing in $J^P=1/2^-$ and $3/2^-$. The effects are again small in $J^P=3/2^-$ but they are sizable for $J^P=1/2^-$. Indeed, the peak around 1780 MeV is shifted to lower energies and becomes considerably broader. This fact is also most welcome since it provides an explanation on why the width of the $1/2^-$ state is bigger than the corresponding state with $3/2^-$, which is supported by experiment, although the masses of the particles in this high energy region are not well determined. We will come to this issue in the next section. We also observe that the second peak around 1900 MeV is shifted to lower energies and widened for $J^P=1/2^-$. The third peak is also widened for $1/2^-$ but there is not much change in its position.

%\subsection{I=0, S=-1}
Fig.~\ref{res3} contains the results of $|T|^2$ for the quantum numbers S=-1, I=1. The first peak appears in 1830 MeV for the channels $\bar{K}^* N$, $\rho \Lambda$, $\rho \Sigma$, $\omega \Sigma$, $\phi \Sigma$, but not for $K^* \Xi$. 
With the only consideration of the vector-baryon channel there was a smooth peak around 2000 MeV, visible in the $\rho \Sigma$ and $K^* \Xi$ channels. What we observe here is that the introduction of the pseudoscalar-baryon channels removes the peak in the $\rho \Sigma$ channel and shifts it to a larger energies around 2180 MeV in the $K^* \Xi$ one.
Once again we see a stronger widening of the first peak for $J^P=1/2^-$ and a slight shift to lower energies with respect to the peak for $J^P=3/2^-$.

In Fig.~\ref{res4} we include the results of S=-2, I=1/2. The results show two peaks very close to each other, one around 2040 MeV and the other one close to 2080 MeV. 
After the introduction of the PB-VB mixing we observe different features in $J^P=1/2^-$ and $3/2^-$. Indeed, the two peaks are still visible for $J^P=3/2^-$ in the $\bar{K}^* \Lambda$ channels, but they separate in the other channels, where only one of them appears in each case. The higher energy resonance shows up in the $\bar{K}^* \Sigma$, $\omega \Xi$ and $\phi \Xi$ channels, and the lower energy resonance shows up only in the $\rho \Xi$ channel.
For $J^P=1/2^-$ the broadening due to the PB-VB mixing removes the two peaks in the $\bar{K}^* \Lambda$ channel and only a broad bump remains, while the two resonances are still visible in the same channels discussed above but the resonances become wider. In the $\bar{K}^* \Sigma$ channel the first peak is also a bit shifted to lower energies.

Using the results of the poles found in this work, we proceed to evaluate the couplings of the different channels to the resonances studied. Those couplings can be found in Tables \ref{tab:couplings12} and \ref{tab:couplings32}, for the resonances found in the scattering matrix.
These couplings give us an idea of which are the most important building blocks in each resonance.

\begin{table}
\begin{center}
\begin{tabular}{cc|cc|cc|cc}
\hline
\multicolumn{8}{c}{Couplings for $J^P=1/2^-$} \\
\hline
S&I&		\multicolumn{2}{c|}{ }				&\multicolumn{2}{c|}{ }&&\\
0 & 1/2 &	\multicolumn{2}{c|}{1690+i24*}				&\multicolumn{2}{c|}{1976+i59}&&\\
\hline
\multicolumn{2}{c|}{Channels} 		& $g_i$ 	& $|g_i|$ 	& $g_i$ & $|g_i|$\\
\hline
\multicolumn{2}{c|}{$\rho N$} 		& 3.1-i1.0	&3.2		&-0.4-i0.4 	&0.6\\
\multicolumn{2}{c|}{$\omega N$} 	& 0.1-i0.2	&0.2		&-1.0-i0.4 	&1.1\\
\multicolumn{2}{c|}{$\phi N$} 		&-0.1+i0.3	&0.3		& 1.5+i0.4 	&1.5\\
\multicolumn{2}{c|}{$K^* \Lambda$}	& 1.8-i1.3 	&2.2		& 2.1-i1.1 	&2.4\\
\multicolumn{2}{c|}{$K^* \Sigma$}	&-0.4-i0.3 	&0.5		& 4.0+i0.0 	&4.0\\
\hline\hline
S&I&		\multicolumn{2}{c|}{ }		&\multicolumn{2}{c|}{ }	&\multicolumn{2}{c}{ }\\
-1 & 0 &	\multicolumn{2}{c|}{1776+i39}	&\multicolumn{2}{c|}{1906+i34*}	&\multicolumn{2}{c}{2163+i37}\\
\hline
\multicolumn{2}{c|}{Channels} 		& $g_i$ & $|g_i|$	& $g_i$ & $|g_i|$	& $g_i$ & $|g_i|$ \\
\hline
\multicolumn{2}{c|}{$\bar{K}^* N$}	& 3.7-i1.3 	&4.0	& 0.1 +i0.3	&0.3	& 0.1 +i0.4 	&0.4\\
\multicolumn{2}{c|}{$\omega\Lambda$}& 1.5-i0.1	&1.5	& 0.4 +i0.2	&0.4	&-0.3 -i0.2  	&0.3\\
\multicolumn{2}{c|}{$\rho \Sigma$} 	&-1.3-i0.2 	&1.3	& 3.6 -i0.9	&3.7	& 0.0 -i0.1 	&0.1\\
\multicolumn{2}{c|}{$\phi \Lambda$}	&-2.0-i0.2 	&2.0	&-0.5 -i0.5 &0.7	& 0.4 +i0.2  	&0.4\\
\multicolumn{2}{c|}{$K^* \Xi$} 		& 0.2-i0.0	&0.2	& 0.3 +i0.2 &0.4	& 3.4 -i0.5 	&3.5\\
\hline\hline
S&I&		\multicolumn{2}{c|}{ } &\multicolumn{2}{c|}{ }&\\
-1 & 1 &	\multicolumn{2}{c|}{1829*}	&\multicolumn{2}{c|}{2116*}&&\\
\hline
\multicolumn{2}{c|}{Channels} 		& $g_i$ 		& $|g_i|$ 	& $g_i$ 		& $|g_i|$& \\
\hline
\multicolumn{2}{c|}{$\bar{K}^* N$} 	& 2.8 +i0.0 	& 2.8		&-0.2 +i 0.2	& 0.3\\
\multicolumn{2}{c|}{$\rho \Lambda$} &-2.4 +i0.2 	& 2.4		&-0.3 +i 0.4	& 0.5\\
\multicolumn{2}{c|}{$\rho \Sigma$} 	&-2.2 +i0.1 	& 2.2		& 2.5 +i 0.0	& 2.5\\
\multicolumn{2}{c|}{$\omega \Sigma$}&-1.3 +i0.1 	& 1.3		&-0.2 +i 0.2	& 0.3\\
\multicolumn{2}{c|}{$K^* \Xi$} 		& 0.2 +i0.1 	& 0.2		& 2.1 -i 0.2	& 2.1\\
\multicolumn{2}{c|}{$\phi \Sigma$}	& 1.8 -i0.2		& 1.8		& 0.2 -i 0.3	& 0.4\\
\hline\hline
S&I&		\multicolumn{2}{c|}{ }		&\multicolumn{2}{c|}{ }&\\
-2 & 1/2 &	\multicolumn{2}{c|}{2047+i19*}	&\multicolumn{2}{c|}{2084*}&&\\
\hline
\multicolumn{2}{c|}{Channels} 				& $g_i$			& $|g_i|$ 	& $g_i$ 		& $|g_i|$& \\
\hline
\multicolumn{2}{c|}{$\bar{K}^* \Lambda$}	&-1.0-i0.2  & 1.0		&-0.1-i0.4 & 0.5\\
\multicolumn{2}{c|}{$\bar{K}^* \Sigma$} 	&-1.3+i0.2	& 1.3		& 2.9+i0.0 & 2.9\\
\multicolumn{2}{c|}{$\rho \Xi$} 			& 2.9-i0.1	& 2.9		& 0.0+i0.8 & 0.8\\
\multicolumn{2}{c|}{$\omega \Xi$} 			& 0.4-i0.3	& 0.5		& 1.5+i0.3 & 1.6\\
\multicolumn{2}{c|}{$\phi \Xi$} 			&-0.6+i0.4	& 0.7		&-2.1-i0.4 & 2.1\\
\hline\hline
\end{tabular}
\caption{Couplings constants for the different channels of the resonances found with $J^P=1/2^-$.}
\label{tab:couplings12}
\end{center}
\end{table}

\begin{table}
\begin{center}
\begin{tabular}{cc|cc|cc|cc}
\hline
\multicolumn{8}{c}{Couplings for $J^P=3/2^-$} \\
\hline
S&I&		\multicolumn{2}{c|}{ }				&\multicolumn{2}{c|}{ }&&\\
0 & 1/2 &	\multicolumn{2}{c|}{1703+i4*}				&\multicolumn{2}{c|}{1979 +i56}&&\\
\hline
\multicolumn{2}{c|}{Channels} 		& $g_i$ 	& $|g_i|$ 	& $g_i$ & $|g_i|$\\
\hline
\multicolumn{2}{c|}{$\rho N$} 		& 2.1-i0.3	&2.1		&-0.4-i0.5 	&0.6\\
\multicolumn{2}{c|}{$\omega N$} 	& 0.1-i0.1	&0.1		&-1.1-i0.4 	&1.2\\
\multicolumn{2}{c|}{$\phi N$} 		&-0.1+i0.1	&0.1		& 1.5+i0.5 	&1.6\\
\multicolumn{2}{c|}{$K^* \Lambda$}	& 1.5-i0.3 	&1.5		& 2.1-i1.0 	&2.4\\
\multicolumn{2}{c|}{$K^* \Sigma$}	&-0.4-i0.0 	&0.4		& 4.0+i0.1 	&4.0\\
\hline\hline
S&I&		\multicolumn{2}{c|}{ }		&\multicolumn{2}{c|}{ }	&\multicolumn{2}{c}{ }\\
-1 & 0 &	\multicolumn{2}{c|}{1786+i11}	&\multicolumn{2}{c|}{1916+i13*}	&\multicolumn{2}{c}{2161+i17}\\
\hline
\multicolumn{2}{c|}{Channels} 		& $g_i$ & $|g_i|$	& $g_i$ & $|g_i|$	& $g_i$ & $|g_i|$ \\
\hline
\multicolumn{2}{c|}{$\bar{K}^* N$}	& 3.4-i0.3 	&3.4	& 0.1 +i0.3	&0.4	& 0.2+i0.3 	&0.4\\
\multicolumn{2}{c|}{$\omega\Lambda$}& 1.4-i0.0	&1.4	& 0.5 +i0.2	&0.5	&-0.3-i0.2  &0.4\\
\multicolumn{2}{c|}{$\rho \Sigma$} 	&-1.3-i0.0 	&1.3	& 3.3 -i0.3	&3.3	& 0.0-i0.1 	&0.1\\
\multicolumn{2}{c|}{$\phi \Lambda$}	&-1.9-i0.0 	&1.9	&-0.7 -i0.3 &0.7	& 0.5+i0.2  &0.5\\
\multicolumn{2}{c|}{$K^* \Xi$} 		& 0.2-i0.0	&0.2	& 0.4 +i0.1 &0.4	& 3.4-i0.5 	&3.4\\
\hline\hline
S&I&		\multicolumn{2}{c|}{ } &\multicolumn{2}{c|}{ }&\\
-1 & 1 &	\multicolumn{2}{c|}{1839*}	&\multicolumn{2}{c|}{2081*}&&\\
\hline
\multicolumn{2}{c|}{Channels} 		& $g_i$ 		& $|g_i|$ 	& $g_i$ 		& $|g_i|$& \\
\hline
\multicolumn{2}{c|}{$\bar{K}^* N$} 	& 2.4 +i0.0 	& 2.4		&-0.2 -i 0.5	& 0.5\\
\multicolumn{2}{c|}{$\rho \Lambda$} &-2.0 +i0.2 	& 2.0		&-0.4 +i 0.7	& 0.8\\
\multicolumn{2}{c|}{$\rho \Sigma$} 	&-1.9 +i0.1 	& 1.9		& 3.2 +i 0.0	& 3.2\\
\multicolumn{2}{c|}{$\omega \Sigma$}&-1.1 +i0.1 	& 1.1		&-0.2 +i 0.4	& 0.4\\
\multicolumn{2}{c|}{$K^* \Xi$} 		& 0.2 +i0.1 	& 0.2		& 2.7 -i 0.4	& 2.7\\
\multicolumn{2}{c|}{$\phi \Sigma$}	& 1.5 -i0.2		& 1.5		& 0.3 -i 0.5	& 0.6\\
\hline\hline
S&I&		\multicolumn{2}{c|}{ }		&\multicolumn{2}{c|}{ }&\\
-2 & 1/2 &	\multicolumn{2}{c|}{2044+i12*}	&\multicolumn{2}{c|}{2082+i5*}&&\\
\hline
\multicolumn{2}{c|}{Channels} 				& $g_i$			& $|g_i|$ 	& $g_i$ 		& $|g_i|$& \\
\hline
\multicolumn{2}{c|}{$\bar{K}^* \Lambda$}	&-1.0-i0.2  & 1.1		&-0.3-i0.2 & 0.4\\
\multicolumn{2}{c|}{$\bar{K}^* \Sigma$} 	&-1.3-i0.1	& 1.3		& 1.6-i0.4 & 1.7\\
\multicolumn{2}{c|}{$\rho \Xi$} 			& 2.9-i0.1	& 2.9		& 0.5+i0.1 & 0.6\\
\multicolumn{2}{c|}{$\omega \Xi$} 			& 0.4-i0.3	& 0.5		& 1.1-i0.1 & 1.1\\
\multicolumn{2}{c|}{$\phi \Xi$} 			&-0.6+i0.4	& 0.7		&-1.4-i0.2 & 1.5\\
\hline\hline
\end{tabular}
\caption{Couplings constants for the different channels of the resonances found with $J^P=3/2^-$.}
\label{tab:couplings32}
\end{center}
\end{table}

\section{Comparison to data}
\begin{table}[ht]

\begin{center}
\begin{tabular}{c|c|cc|ccccc}\hline\hline
$I,\,S$&\multicolumn{3}{c|}{Theory} & \multicolumn{5}{c}{PDG data}\\
\hline
%    \vspace*{-0.5cm}
    & pole position    & \multicolumn{2}{c|}{real axis} &  &  & &  &  \\

    &   & mass & width &name & $J^P$ & status & mass & width \\
    \hline
$1/2,0$ & --- & 1696  & 92  & $N(1650)$ & $1/2^-$ & $\star\star\star\star$ & 1645-1670
& 145-185\\
  &      &       &     & $N(1700)$ & $3/2^-$ & $\star\star\star$ &
	1650-1750 & 50-150\\

       & $1977 + {\rm i} 53$  & 1972  & 64  & $N(2080)$ & $3/2^-$ & $\star\star$ & $\approx 2080$
& 180-450 \\	
   &     &       &     & $N(2090)$ & $1/2^-$ & $\star$ &
 $\approx 2090$ & 100-400 \\
 \hline
$0,-1$ & $1784 + {\rm i} 4$ & 1783  & 9  & $\Lambda(1690)$ & $3/2^-$ & $\star\star\star\star$ &
1685-1695 & 50-70 \\
  &       &       &    & $\Lambda(1800)$ & $1/2^-$ & $\star\star\star$ &
1720-1850 & 200-400 \\
       & $1907 + {\rm i} 70$ & 1900  & 54  & $\Lambda(2000)$ & $?^?$ & $\star$ & $\approx 2000$
& 73-240\\
       & $2158 + {\rm i} 13$ & 2158  & 23  &  &  &  & & \\
       \hline
$1,-1$ &  ---  & 1830  & 42  & $\Sigma(1750)$ & $1/2^-$ & $\star\star\star$ &
1730-1800 & 60-160 \\
  & ---    & 1987  & 240  & $\Sigma(1940)$ & $3/2^-$ & $\star\star\star$ & 1900-1950
& 150-300\\
   &     &       &   & $\Sigma(2000)$ & $1/2^-$ & $\star$ &
$\approx 2000$ & 100-450 \\\hline
$1/2,-2$ & $2039 + {\rm i} 67$ & 2039  & 64  & $\Xi(1950)$ & $?^?$ & $\star\star\star$ &
$1950\pm15$ & $60\pm 20$ \\
         & $2083 + {\rm i} 31 $ &  2077     & 29  &  $\Xi(2120)$ & $?^?$ & $\star$ &
$\approx 2120$ & 25  \\
 \hline\hline
    \end{tabular}
\caption{The properties of the 9 dynamically generated resonances and their possible PDG
counterparts.}
\label{tab:pdg}
\end{center}
\end{table}

In Tables \ref{tab:pdg12} and \ref{tab:pdg32} we show a summary of the results obtained and the tentative association to known states of the PDG \cite{pdg}. In Table \ref{tab:pdg12} we show the states for $J^P=1/2^-$ and in Table \ref{tab:pdg32} for $J^P=3/2^-$. For comparison, the results of Ref.~\cite{angelsvec} without the mixing are displayed in Table \ref{tab:pdg}.

\begin{table}[ht]
\begin{center}
\begin{tabular}{c|c|cc|ccccc}
\hline\hline
$S,\,I$	&\multicolumn{3}{c|}{Theory} & \multicolumn{5}{c}{PDG data}\\
\hline
    	& pole position	& \multicolumn{2}{c|}{real axis} &  &  & &  &  \\
    	& $M_R+i\Gamma /2$	& mass & width &name & $J^P$ & status & mass & width \\
\hline
$0,1/2$ & $1690+i24^{*}			$	& 1658  & 98  
		& $N(1650)$ & $1/2^-$ & $\star\star\star\star$ 	& 1645-1670	& 145-185\\
  		
       	& $1979+i67			$	& 1973 & 85 
   		& $N(2090)$ & $1/2^-$ & $\star$ 	 & $\approx 2090$ & 100-400 \\
\hline
$-1,0$	& $1776+i39				$	& 1747 & 94  
  		& $\Lambda(1800)$ & $1/2^-$ & $\star\star\star$ 		& 1720-1850 & 200-400 \\
  		
        & $1906+i34^{*}			$ 	& 1890 & 93  
        & $\Lambda(2000)$ & $?^?$ & $\star$ 					& $\approx 2000$ & 73-240\\

        & $2163+i37				$ 	& 2149 & 61  &  &  &  & & \\
\hline
$-1,1$  & $ -			$	& 1829& 84  
		& $\Sigma(1750)$ & $1/2^-$ & $\star\star\star$ & 1730-1800 & 60-160 \\
   		& $ -			 $ 	& 2116 & 200-240  
   		& $\Sigma(2000)$ & $1/2^-$ & $\star$ 			& $\approx 2000$ & 100-450 \\
\hline
$-2,1/2$& $2047+i19^{*}	$	& 2039 & 70  
		& $\Xi(1950)$ & $?^?$ & $\star\star\star$ 	& $1950\pm15$ & $60\pm 20$ \\

        & $ -			$	& 2084 & 53   
        & $\Xi(2120)$ & $?^?$ & $\star$ 			& $\approx 2120$ & 25  \\
\hline\hline
\end{tabular}
\caption{The properties of the nine dynamically generated resonances and their possible PDG
counterparts for $J^P=1/2^-$. The numbers with asterisk in the imaginary part of the pole position are obtained without the convolution for the vector mass distribution of the $\rho$ and $K^*$.}
\label{tab:pdg12}
\end{center}
\end{table}

\begin{table}[ht]
\begin{center}
\begin{tabular}{c|c|cc|ccccc}
\hline\hline
$S,\,I$	&\multicolumn{3}{c|}{Theory} & \multicolumn{5}{c}{PDG data}\\
\hline
    	& pole position	& \multicolumn{2}{c|}{real axis} &  &  & &  &  \\
    	& $M_R+i\Gamma /2$	& mass & width &name & $J^P$ & status & mass & width \\
\hline
$0,1/2$ & $1703+i4^{*}			$	& 1705  & 103  
  		& $N(1700)$ & $3/2^-$ & $\star\star\star$ 		& 1650-1750 & 50-150\\
  		
       	& $1979+i56			$	& 1975 & 72 
       	& $N(2080)$ & $3/2^-$ & $\star\star$ & $\approx 2080$ & 180-450 \\	
\hline
$-1,0$	& $1786+i11				$	& 1785 & 19  
		& $\Lambda(1690)$ & $3/2^-$ & $\star\star\star \star$ 	& 1685-1695 & 50-70 \\
  		
        & $1916+i13^{*}			$ 	& 1914 & 59  
        & $\Lambda(2000)$ & $?^?$ & $\star$ 					& $\approx 2000$ & 73-240\\

        & $2161+i17				$ 	& 2158 & 29  &  &  &  & & \\
\hline
$-1,1$  & $ -			$	& 1839& 58  
  		& $\Sigma(1940)$ & $3/2^-$ & $\star\star\star$ & 1900-1950 & 150-300\\
   		& $ -			 $ 	& 2081 & 270  &  &  &  & & \\  
\hline
$-2,1/2$& $2044+i12^{*}	$	& 2040 & 53  
		& $\Xi(1950)$ & $?^?$ & $\star\star\star$ 	& $1950\pm15$ & $60\pm 20$ \\

        & $2082+i5^{*} 	$	& 2082 & 32   
        & $\Xi(2120)$ & $?^?$ & $\star$ 			& $\approx 2120$ & 25  \\
\hline\hline
\end{tabular}
\caption{The properties of the nine dynamically generated resonances and their possible PDG
counterparts for $J^P=3/2^-$. The numbers with asterisk in the imaginary part of the pole position are obtained without the convolution for the vector mass distribution of the $\rho$ and $K^*$.}
\label{tab:pdg32}
\end{center}
\end{table}

For S=0, I=1/2 we find a state around 1658 MeV with $J^P=1/2^-$. With fixed $\rho$ mass the peak has a pole associated but with a small imaginary part. Yet, the consideration of the convolution the $\rho$ mass and the pseudoscalar-baryon channels widens the structure considerably becoming an approximate Breit-Wigner structure with a width of about 98 MeV. This width is compatible with the values of the $N^*(1650)~(1/2^-)$ to which the peak obtained would be associated.
A similar behavior  is seen for $3/2^-$ and we find a state at 1705 MeV which can be associated to the $N^*(1700)~(3/2^-).$
In the case of the second peak, the mass found is around 1975 MeV both for $1/2^-$ and $3/2^-$, a bit smaller than the nominal experimental masses of the $N^*(2080)~(3/2^-)$ and $N^*(2090)~(1/2^-)$ resonances cataloged in the PDG \cite{pdg}. 
However, we should note that the masses associated in the PDG are averages done there, but there is a large dispersion of the data for the masses and our calculated results fit well within the experimental masses. The width is also compatible with the experimental results within the large experimental range.

In the case of S=-1 and I=0, we found three peaks. A resonance found at 1786 MeV for $3/2^-$ could be associated to $\Lambda (1690)$ with $J^P=3/2^-$ and the one at 1747 MeV with $1/2^-$ could be associated to the $\Lambda (1800)$ with $J^P=1/2^-$.
Once again, there is a large variation for the masses in the different experiments reported in the PDG under the umbrella of the $\Lambda (1800)$ and the mass that we obtain fits well within these values. For the case of the $\Lambda (1690)$ the dispersion of the masses is much smaller, with the values around 1690 MeV, smaller than our calculated result. The relatively small width of 40 MeV that we obtain is compatible to the experimental widths. For the case of the $\Lambda (1800)$ there is a large variation of the widths, with values as low as 40 or 100 MeV.
We also found a state at 1890 MeV for $1/2^-$ and another at 1914 MeV for $3/2^-$ and widths 93 MeV and 59 MeV respectively. They could be associated to the $\Lambda (2000)$ which has an unknown spin in PDG, but the mass and the width calculated are compatible with this state.
A third state is found with a mass around 2150 MeV. More concretely, there is the $J^P=1/2^-$ at 2149 MeV and width 61 MeV and the $J^P=3/2^-$ partner at 2158 MeV with a smaller width of 29 MeV. We do not find counterparts in the PDG, so these are predictions of the theory.

For S=-1 and I=1, we find two peaks. The first peak for $1/2^-$ with mass 1829 MeV and width 84 MeV can be associated to the $\Sigma (1750)~(1/2^-)$ and its partner of $3/2^-$ at 1839 and width 58 MeV can be associated to the $\Sigma (1940)~(3/2^-)$. The widths are in good agreement with the experimental results considering the dispersion of data for experiments collected under the same umbrella of this resonance. The second peak has a mass of 2116 MeV and a width of about 200 MeV, and the only state that could correspond to this resonance is $\Sigma (2000)~(1/2^-)$ which has an experimental width compatible with that result.

As one can see in Table~\ref{tab:pdg}, the width provided in Ref.~\cite{angelsvec} for this state is bigger (240 MeV) than the one reported here. This is a surprise since we should expect an increase of the width from the inclusion of the PB decay channels. This is the only case where this happens. However, one can understand the reason. In both cases the width is obtained from a visual inspection of $|T|^2$ in the real axis. However, no background determination is done, so this should be taken only as a rough estimate. Indeed, in Fig.~\ref{res3}, channel $K^* \Xi$ in left panel, the shape obtained in both cases does not resemble a Breit-Wigner and the association of a width to it has been qualitative. In view of that, we have now put a width of 200-240 MeV which is compatible with the one quoted in Ref.~\cite{angelsvec} if a similar error band would have been taken in Ref.~\cite{angelsvec}.

Finally, for the case of S=-2 and I=1/2 we found two peaks. For this case it is not clear to which states one can associate them since in this region the states cataloged in the PDG have no determined spin, but the states $\Xi (1950)$ and $\Xi (2120)$ could be associated to those peaks. 
Both the masses and the widths for both states are compatible with the theoretical results. We should also note that an experimental search devoted to the S=-2 sector is being conducted at Jefferson Lab \cite{Nefkens:2006bc,Price:2004xm}.

\section{Conclusions}

The interaction of vector mesons with the baryon octet using the hidden gauge formalism, produces nine resonances dynamically generated, degenerate in $J^P=1/2^-,~3/2^-$, which can be associated to states of the PDG. However the results show that the theoretical widths are significantly smaller than the experimental ones. So, one could think that there is something else involved in the vector meson - baryon octet interaction. Since, pseudoscalar mesons couple both, to vectors and baryons, one can think that it is possible that the interaction is mediated not only by a vector exchange but also by a pseudoscalar exchange. 

We found a mechanism, the box diagram involving $V\rightarrow PP$ in two vertices, which is common to $J^P=1/2^-$ and $3/2^-$ and gives the same contribution in both cases. In addition to this mechanism, there are other ones involving a contact term (vertex correction of the Kroll-Ruderman type) which only contribute for $J^P=1/2^-$, thus breaking the original degeneracy between these states. The box diagram gives a relatively small contribution, but the terms involving the contact term are more important, producing a widening of the resonance and some times a small shift to smaller energies. In particular we found very rewarding that this splitting leads to a good phenomenological result in the case of the $N^*(1650)(1/2^-)$ and $N^*(1700)(3/2^-)$, which are well reproduced both for the masses and widths. The tendency to have a bigger width in the state of $1/2^-$ than in the partner with $3/2^-$ is also observed in the low lying resonances. For higher mass resonances we noted the difficulty to establish a clean correspondence with particles in the PDG given the large fluctuation of results between different experiments.

The study of the scattering matrix reveals that in fact the widths of the resonances found with this new potential, are bigger than the original ones, and are in a better agreement with the experimental results found in the PDG. Also the couplings are studied, and although the real and imaginary parts are somewhat different from the original ones, the moduli remain very similar for most of the cases.

The formalism done here could be extended to the complementary problem of studying the effect of the vector baryon channels in the resonances which are largely made of pseudoscalar-baryon channels. The interplay of pseudoscalar-baryon, vector-baryon and $\gamma$-baryon channels is emerging also as a new experimental line in different reactions, like the photoproduction of $K^* \Lambda$ close to threshold \cite{hicks} and $K \Lambda$ photoproduction close to the $K^* \Lambda$ threshold \cite{schmieden}. The formalism developed here should be very useful to tackle theoretically these works.

\section{Acknowledgments}
We would like to thank K. Khemchandani for useful discussions.
This work is partly supported by DGICYT contracts  FIS2006-03438, FPA2007-62777, the Generalitat Valenciana in the program Prometeo and  the EU Integrated Infrastructure Initiative Hadron Physics Project  under Grant Agreement n.227431.

\appendix

\section{Expression of the Box diagram integral}
\label{appendix:boxint}
The integral of the box diagram contains a function which comes from the propagator of the three mesons and the baryon of the loop. This expression is manipulated in order to cancel the fallacious poles which appears when one integrate the zero component of the four-momentum using the Cauchy's Theorem. The sum of the residues terms leads to a long expression that is given below. 
\begin{eqnarray}
Den &=& (-2\omega_1 + i\epsilon)(-P^0_V-\omega_1-\omega_2 + i\epsilon) (\alpha - \omega_3 -\omega_1 + i\epsilon) 	\nonumber \\
	& & (\beta - \omega_1 + i\epsilon) (P^0_V-\omega_2-\omega_1 + i\epsilon) (-2\omega_2 + i\epsilon)
	\nonumber \\
	& & (P^0_V+\alpha-\omega_2-\omega_3 + i\epsilon)(\beta + P^0_V- \omega_2 + i\epsilon)(-\alpha-\omega_3-\omega_1 + i\epsilon)
	\nonumber \\
	& & (-2\omega_3 + i\epsilon)(-\alpha-P^0_V-\omega_3-\omega_2 + i\epsilon)(\beta -\alpha - \omega_3 + i\epsilon)
\label{eq:den}
\end{eqnarray}
where 
$$\omega_i=\sqrt{|\vec{q}|^2+m^2_i},~~~~\alpha=P^0_{B_1}-P^0_{B_2},~~~~\beta=P^0_{B_1}-E_B(\vec{q}),$$
and
$$P^0_{B_i}=\frac{s+M^2_{B_i}-M^2_{V_i}}{2\sqrt{s}},~~~~P^0_{V_i}=\frac{s+M^2_{V_i}-M^2_{M_i}}{2\sqrt{s}}, ~~~~E_B(\vec{q})=\sqrt{|\vec{q}|^2+M^2_B}.$$

\begin{eqnarray}
Num &=&  {P^0_V}^3 \omega_2 (\omega_1^2 + \omega_3 (\alpha - \beta + \omega_3) + \omega_1 (-\beta + 2 \omega_3)) 
\nonumber \\
& &		-{P^0_V}^2 \omega_2 ((-2 \alpha - \beta + \omega_2) \omega_3 (\alpha - \beta + \omega_3) 
 		+\omega_1^2 (-\beta + \omega_2 + 2 \omega_3) 
\nonumber \\
& & 	+\omega_1 (\beta^2 - 2 (\alpha + \beta) \omega_3 + 2 \omega_3^2 + \omega_2 (-\beta + 2 \omega_3))) 
\nonumber \\
& &		- P^0_V \omega_2 (\omega_1^4 + \omega_1^3 (-\beta + 2 \omega_2 + 2 \omega_3) + 
            \omega_3 (\alpha - \beta + \omega_3)(-(\alpha (\alpha + 2 \beta)) 
\nonumber \\
& &		+ \omega_2^2 + \omega_3^2 + 2 \omega_2 (\alpha + \omega_3)) + 
            \omega_1 (4 \omega_2 \omega_3 (\alpha - \beta + \omega_3)
\nonumber \\
& &		+ 2 \omega_3 (\alpha + \omega_3) (\alpha - \beta + \omega_3) + \omega_2^2 (-\beta + 2 \omega_3)) + 
            \omega_1^2 (\omega_2^2 + 2 \omega_3 (\alpha - \beta + \omega_3) 
\nonumber \\
& &	 	+ \omega_2 (-2 \beta + 4 \omega_3))) + (\omega_1 + \omega_2) 
			((-\beta + \omega_2) \omega_3 (\alpha - \beta + \omega_3) 
             (-\alpha^2 + \omega_2^2 + 2 \omega_2 \omega_3 + \omega_3^2) 
\nonumber \\
& &     + \omega_1^3 (\omega_2^2 + \omega_3 (\alpha - \beta + \omega_3) + \omega_2 (-\beta + 2 \omega_3)) + 
            \omega_1^2 (-\beta + \omega_2 + 2 \omega_3) (\omega_2^2 
\nonumber \\
& & 	+ \omega_3 (\alpha - \beta + \omega_3) + \omega_2 (-\beta + 2 \omega_3)) + 
            \omega_1 (\omega_2^3 (-\beta + 2 \omega_3)
\nonumber \\
& &  	+ \omega_2 \omega_3 (-2 \alpha^2 - \alpha \beta + 3 \beta^2 + 2 \alpha \omega_3 - 7 \beta \omega_3 + 
                  4 \omega_3^2) + \omega_2^2 (\beta^2 + (\alpha - 5 \beta) \omega_3 + 5 \omega_3^2) 
\nonumber \\
& & 	+ \omega_3 (\alpha^2 (-\alpha + \beta) -  (\alpha^2 + 2 \alpha \beta - 2 \beta^2) 
                   \omega_3 + (\alpha - 3 \beta) \omega_3^2 + \omega_3^3)))
\label{eq:num}
\end{eqnarray}

\section{Matrix elements of the $\vec{\sigma}~ \vec{\epsilon}$ operator}
\label{appendix:spin}

Let us evaluate the matrix element
\begin{equation}
\left\langle 1/2 m' | \vec{\sigma}~ \vec{\epsilon} | J M \right\rangle
\end{equation}
where
\begin{equation}
| J M > = \sum_m {\cal C}(1/2,1,J;m,M-m,M)|1/2,m>|\vec{\epsilon}_{M-m}>
\end{equation}
and, as usual,
\begin{equation}
 \vec{\epsilon}_\mu = 
   \left \{
      \begin{array}{rcl}
          -\frac{1}{\sqrt{2}} (\vec{\epsilon}_1+i \vec{\epsilon}_2) \\
         \vec{\epsilon}_3 \\ 
         \frac{1}{\sqrt{2}} (\vec{\epsilon}_1-i \vec{\epsilon}_2) 
      \end{array}
   \right .
   ~~~~~\mu=1,~0,~-1
\end{equation}
Hence the matrix element can be written as
\begin{eqnarray}
\left\langle 1/2 m' | \vec{\sigma}~ \vec{\epsilon} | J M \right\rangle & = & \sum_m {\cal C}(1/2,1,J;m,M-m,M)
\left\langle 1/2 m'| \vec{\sigma} \vec{\epsilon}_{M-m} |1/2,m \right\rangle \\
 & = & \sum_m {\cal C}(1/2,1,J;m,M-m,M) \left\langle 1/2 m'| \sigma_{M-m} |1/2,m \right\rangle
 \label{eq:matrixelementsigma}
\end{eqnarray}
Using the Wigner-Eckart theorem one obtains that
\begin{equation}
\left\langle 1/2 m'| \sigma_{M-m} |1/2,m \right\rangle = \sqrt{3}~{\cal C}(1/2,1,1/2;m,M-m,m')
\end{equation}
from where we get that $m' \equiv M$. Substituting this result in Eq.~(\ref{eq:matrixelementsigma}) we obtain
\begin{equation}
\sum_m {\cal C}(1/2,1,J;m,M-m,M) \sqrt{3}~{\cal C}(1/2,1,1/2;m,M-m,M)=\sqrt{3}~\delta_{J,1/2}
\end{equation}
Hence, only $J=1/2$ contributes and $m'=M$.

\section{Coefficients of the Baryon octet - pseudoscalar mesons interaction}
\label{appendix:BBPcoeff}
The coefficients of the BBP interaction $\alpha$ and $\beta$ which are related to the BBP vertex as
\begin{equation}
-i t_{BBP} = \left\lbrace \alpha \frac{\left(D+F\right)}{2 f} + \beta \frac{\left(D-F\right)}{2 f}  \right\rbrace \sigma \vec{k}
\label{lbbpcoef}
\end{equation}
are given in Table \ref{tab:BBPcoeff}.

\begin{table}
\begin{footnotesize}
\renewcommand{\arraystretch}{1.3}
\begin{center}
\begin{tabular}{c|c|c|c|c|c|c|c|c}
\hline
\hline
\multicolumn{9}{c}{Coefficients $\alpha$ for $(D+F)/2f$}\\
\hline
\hline
$\eta_8$	&$\bar{n}~n$&$\bar{p}~p$&$\bar{\Sigma}^-~\Sigma^-$&$\bar{\Sigma}^+~\Sigma^+$
 		&$\bar{\Sigma}^0~\Sigma^0$&$\bar{\Lambda}~\Lambda$&$\bar{\Xi}^0~\Xi^0$&$\bar{\Xi}^-~\Xi^-$\\
%\hline
		&$\frac{1}{\sqrt{3}}$&$\frac{1}{\sqrt{3}}$&$\frac{1}{\sqrt{3}}$
		&$\frac{1}{\sqrt{3}}$&$\frac{1}{\sqrt{3}}$&$-\frac{1}{\sqrt{3}}$
		&$-\frac{2}{\sqrt{3}}$&$-\frac{2}{\sqrt{3}}$ \\
\hline
$\pi^0$	&$\bar{n}~n$&$\bar{p}~p$&$\bar{\Sigma}^-~\Sigma^-$&$\bar{\Sigma}^+~\Sigma^+$
 		&$\bar{\Sigma}^0~\Lambda$&$\bar{\Lambda}~\Sigma^0$&& \\
%\hline 
		&$-1$&$1$&$-1$
		&$1$&$\frac{1}{\sqrt{3}}$&$\frac{1}{\sqrt{3}}$&& \\
\hline
$\pi^-$	&$\bar{n}~p$&&$\bar{\Sigma}^0~\Sigma^+$&$\bar{\Sigma}^-~\Sigma^0$
 		&$\bar{\Lambda}~\Sigma^+$&$\bar{\Sigma}^-~\Lambda$&& \\
%\hline 
		&$\sqrt{2}$&&$-1$&$1$
		&$\frac{1}{\sqrt{3}}$&$\frac{1}{\sqrt{3}}$&& \\
\hline
$\pi^+$	&$\bar{p}~n$&&$\bar{\Sigma}^+~\Sigma^0$&$\bar{\Sigma}^0~\Sigma^-$
 		&$\bar{\Lambda}~\Sigma^-$&$\bar{\Sigma}^+~\Lambda$&& \\
%\hline 
		&$\sqrt{2}$&&$-1$&$1$
		&$\frac{1}{\sqrt{3}}$&$\frac{1}{\sqrt{3}}$&& \\
\hline
\hline
$K^-$	&$\bar{\Lambda}~p$&$\bar{\Xi}^0~\Sigma^+$
 		&$\bar{\Xi}^-~\Sigma^0$&$\bar{\Xi}^-~\Lambda$&&&& \\
%\hline 
		&$-\frac{2}{\sqrt{3}}$&$\sqrt{2}$
		&$1$&$\frac{1}{\sqrt{3}}$&&&& \\
\hline
$K^+$	&$\bar{p}~\Lambda$&$\bar{\Sigma}^+~\Xi^0$&$\bar{\Sigma}^0~\Xi^-$
 		&$\bar{\Lambda}~\Xi^-$&&&& \\
%\hline 
		&$-\frac{2}{\sqrt{3}}$&$\sqrt{2}$
		&$1$&$\frac{1}{\sqrt{3}}$&&&& \\
\hline
$\bar{K}^0$	&$\bar{\Lambda}~n$&$\bar{\Xi}^-~\Sigma^-$
 		&$\bar{\Xi}^0~\Sigma^0$&$\bar{\Xi}^0~\Lambda$&&&& \\
%\hline 
		&$-\frac{2}{\sqrt{3}}$&$\sqrt{2}$
		&$-1$&$\frac{1}{\sqrt{3}}$&&&& \\
\hline
$K^0$	&$\bar{n}~\Lambda$&$\bar{\Sigma}^-~\Xi^-$&$\bar{\Sigma}^0~\Xi^0$
 		&$\bar{\Lambda}~\Xi^0$&&&& \\
%\hline 
		&$-\frac{2}{\sqrt{3}}$&$\sqrt{2}$
		&$-1$&$\frac{1}{\sqrt{3}}$&&&& \\	
\hline
\hline
\multicolumn{9}{c}{Coefficients $\beta$ for $(D-F)/2f$}\\
\hline
\hline
$\eta_8$&$\bar{n}~n$&$\bar{p}~p$&$\bar{\Sigma}^-~\Sigma^-$&$\bar{\Sigma}^+~\Sigma^+$
 		&$\bar{\Sigma}^0~\Sigma^0$&$\bar{\Lambda}~\Lambda$&$\bar{\Xi}^-~\Xi^-$&$\bar{\Xi}^0~\Xi^0$\\
%\hline
		&$-\frac{2}{\sqrt{3}}$&$-\frac{2}{\sqrt{3}}$&$\frac{1}{\sqrt{3}}$
		&$\frac{1}{\sqrt{3}}$&$\frac{1}{\sqrt{3}}$&$-\frac{1}{\sqrt{3}}$
		&$\frac{1}{\sqrt{3}}$&$\frac{1}{\sqrt{3}}$ \\
\hline
$\pi^0$	&&&$\bar{\Sigma}^-~\Sigma^-$&$\bar{\Sigma}^+~\Sigma^+$
 		&$\bar{\Sigma}^0~\Lambda$&$\bar{\Lambda}~\Sigma^0$&$\bar{\Xi}^-~\Xi^-$&$\bar{\Xi}^0~\Xi^0$ \\
%\hline 
		&&&$1$&$-1$
		&$\frac{1}{\sqrt{3}}$&$\frac{1}{\sqrt{3}}$&$1$&$-1$ \\
\hline
$\pi^-$	&&&$\bar{\Sigma}^-~\Sigma^0$&$\bar{\Sigma}^0~\Sigma^+$
 		&$\bar{\Sigma}^-~\Lambda$&$\bar{\Lambda}~\Sigma^+$&$\bar{\Xi}^-~\Xi^0$& \\
%\hline 
		&&&$-1$&$1$
		&$\frac{1}{\sqrt{3}}$&$\frac{1}{\sqrt{3}}$&$\sqrt{2}$& \\
\hline
$\pi^+$	&&&$\bar{\Sigma}^0~\Sigma^-$&$\bar{\Sigma}^+~\Sigma^0$
 		&$\bar{\Sigma}^+~\Lambda$&$\bar{\Lambda}~\Sigma^-$&$\bar{\Xi}^0~\Xi^-$& \\
%\hline 
		&&&$-1$&$1$
		&$\frac{1}{\sqrt{3}}$&$\frac{1}{\sqrt{3}}$&$\sqrt{2}$& \\
\hline
\hline
$K^-$	&$\bar{\Lambda}~p$&$\bar{\Sigma}^0~p$&$\bar{\Sigma}^-~n$&$\bar{\Xi}^-~\Lambda$&&&& \\
%\hline 
		&$\frac{1}{\sqrt{3}}$&$1$&$\sqrt{2}$&$-\frac{2}{\sqrt{3}}$&&&& \\
\hline
$K^+$	&$\bar{p}~\Lambda$&$\bar{p}~\Sigma^0$&$\bar{n}~\Sigma^-$&$\bar{\Lambda}~\Xi^-$&&&& \\
%\hline 
		&$\frac{1}{\sqrt{3}}$&$1$&$\sqrt{2}$&$-\frac{2}{\sqrt{3}}$&&&& \\
\hline
$\bar{K}^0$	&$\bar{\Lambda}~n$&$\bar{\Sigma}^0~n$&$\bar{\Sigma}^+~p$&$\bar{\Xi}^0~\Lambda$&&&& \\
%\hline 
		&$\frac{1}{\sqrt{3}}$&$-1$&$\sqrt{2}$&$-\frac{2}{\sqrt{3}}$&&&& \\
\hline
$K^0$	&$\bar{n}~\Lambda$&$\bar{n}~\Sigma^0$&$\bar{p}~\Sigma^+$&$\bar{\Lambda}~\Xi^0$&&&& \\
%\hline 
		&$\frac{1}{\sqrt{3}}$&$-1$&$\sqrt{2}$&$-\frac{2}{\sqrt{3}}$&&&& \\	
\hline

\end{tabular}
\end{center}
\caption{Coefficients for the BBP vertex.}
\label{tab:BBPcoeff}
\end{footnotesize}
\end{table}

\section{Coefficients of the box integral}
\label{appendix:boxintcoeff}
The box diagram generates an integral that we have analytically calculated in the general case in Eq. (\ref{boxint}). The expression has four coefficients ($a$, $b$, $c$ and $C_{V_1}C_{V_2}$) which depend on every channel and particles involved in the loop, the baryon $B$ and the three pseudoscalar mesons ($B_1$, $B_2$ and $B_3$ ). The Tables \ref{tab:intboxcoeff1}, \ref{tab:intboxcoeff2}, \ref{tab:intboxcoeff3} and \ref{tab:intboxcoeff4}, contain those coefficients for each quantum number, channel and loop.
\begin{table}
\begin{footnotesize}
\renewcommand{\arraystretch}{1.2}
\begin{center}
\begin{tabular}{c|c|c|c|c|c}
\hline
\multicolumn{6}{c}{Coefficients for state $S=0,~I=1/2$}\\
\hline
\hline
Channel	&Box	&$(D+F)^2$&	$(D-F)^2$	&$(D+F)(D-F)$	&$C_{V_1}C_{V_2}$ \\
		&$B P_1 P_2 P_3$ & a & b & c & \\
\hline
$\rho N$	&$N\pi\pi\pi$	&1				&0				&0				&$\frac{4}{3}$ \\
			&$\Sigma KKK$	&0				&$\frac{1}{4}$	&0				&1 \\
			&$\Lambda KKK$	&1				&$\frac{1}{4}$	&-1				&1 \\
\hline
$\omega N$	&$\Sigma KKK$	&0				&1				&0				&$\frac{3}{4}$	\\
			&$\Lambda KKK$	&1				&$\frac{1}{4}$	&-1				&$\frac{1}{3}$	\\
\hline
$\phi N$	&$\Sigma KKK$	&0				&1				&0				&$\frac{3}{2}$	\\
			&$\Lambda KKK$	&1				&$\frac{1}{4}$	&-1				&$\frac{2}{3}$	\\
\hline
$K^*\Lambda$&$N K \pi K$	&1				&$\frac{1}{4}$	&-1				&1 \\
			&$\Sigma\pi K\pi$&1				&1				&2				&$\frac{1}{4}$ \\
			&$N \eta K\eta$	&1				&1				&2				&$\frac{1}{4}$ \\
			&$N K \eta K$	&1				&$\frac{1}{4}$	&-1				&1 \\
\hline
$K^*\Sigma$	&$N K \pi K$	&0				&1				&0				&1 \\
			&$\Sigma\pi K\pi$&1				&1				&-2				&$\frac{11}{12}$ \\
			&$\Sigma\eta K\eta$&1			&1				&2				&$\frac{1}{4}$ \\
			&$\Lambda\eta K\eta$&1			&1				&2				&$\frac{1}{18}$ \\
			&$\Sigma\pi K\eta$&1			&-1				&0				&-$\frac{1}{3}$ \\
			&$\Sigma\eta K\pi$&1			&-1				&0				&$\frac{1}{3}$ \\
\hline
\hline
\multicolumn{6}{c}{Non-diagonal terms}\\
\hline
\hline
Channel	&Box	&$(D+F)^2$&	$(D-F)^2$	&$(D+F)(D-F)$	&$C_{V_1}C_{V_2}$ \\
		&$B P_1 P_2 P_3$ & a & b & c & \\
\hline
$\rho N \rightarrow K^* \Lambda$
			&$\Lambda KK\eta$	&2	&-1	& 1				&$\frac{1}{4}$ \\
			&$\Sigma KK\pi$		&0	& 1	& 1				&-$\frac{1}{4}$ \\
			&$N \pi \pi K$		&2	& 0	&-1				&1 \\
\hline
$\rho N \rightarrow K^* \Sigma$
			&$\Lambda KK\pi$	&2	&-1	& 1				&$\frac{1}{4}$ \\
			&$\Sigma KK\eta$	&0	& 1	& 1				&-$\frac{1}{4}$ \\
			&$\Sigma KK\pi$		&0	&-1	& 1				&-$\frac{1}{2}$ \\
			&$N \pi \pi K$		&0	& 0	& 1				&$\frac{5}{3}$ \\
\hline
$K^* \Lambda \rightarrow K^* \Sigma$
			&$\Sigma \pi K\eta$		&1	& 1	& 2				&$\frac{1}{4}$ \\
			&$\Sigma \pi K\pi$		&1	&-1	& 0				&$\frac{1}{3}$ \\
			&$\Lambda \eta K\pi$	&1	& 1	& 2				&-$\frac{1}{4}$ \\
			&$N K \pi K$			&0	&-1	& 2				&$\frac{5}{12}$ \\
			&$N K \eta K$			&0	&-1	& 2				&-$\frac{1}{4}$ \\
\hline
\hline
\end{tabular}
\end{center}
\caption{Coefficients of the Box Integral for the S=0, I=1/2 states.}
\label{tab:intboxcoeff1}
\end{footnotesize}
\end{table}

\begin{table}
\begin{footnotesize}
\renewcommand{\arraystretch}{1.2}
\begin{center}
\begin{tabular}{c|c|c|c|c|c}
\hline
\multicolumn{6}{c}{Coefficients for state $S=-1,~I=0$}\\
\hline
\hline
Channel	&Box	&$(D+F)^2$&	$(D-F)^2$	&$(D+F)(D-F)$	&$C_{V_1}C_{V_2}$ \\
		&$B P_1 P_2 P_3$ & a & b & c & \\
\hline
$\bar{K}^* N$	&$\Sigma K \pi K$	&0				&1				&0				&$\frac{3}{2}$ \\
				&$N \pi K \pi$		&1				&0				&0				&1 \\
				&$\Lambda K \eta K$	&1				&$\frac{1}{4}$	&-1				&2 \\
				&$N \eta K \eta$	&$\frac{1}{4}$	&1				&-1				&2 \\
\hline
$\omega \Lambda$&$\Xi KKK$			&$\frac{1}{4}$	&1				&-1				&$\frac{2}{3}$	\\
				&$N KKK$			&1				&$\frac{1}{4}$	&-1				&$\frac{2}{3}$	\\
\hline
$\rho \Sigma$&$\Sigma \pi\pi\pi$	&1				&1				&-2				&$\frac{10}{3}$ \\
			&$\Xi KKK$				&1				&0				&0				&$\frac{5}{6}$	\\
			&$N KKK$				&0				&1				&0				&$\frac{5}{6}$	\\
\hline
$\phi \Lambda$	&$\Xi KKK$			&$\frac{1}{4}$	&1				&-1				&$\frac{4}{3}$	\\
				&$N KKK$			&1				&$\frac{1}{4}$	&-1				&$\frac{4}{3}$	\\
\hline
$K^*\Xi$	&$\Sigma K \pi K$		&1				&0				&0				&1 \\
			&$\Xi \pi K\pi$			&0				&1				&0				&$\frac{5}{4}$ \\
			&$\Lambda K \pi K$		&$\frac{1}{4}$	&1				&-1				&$\frac{1}{3}$ \\
			&$\Xi \eta K\eta$		&1				&$\frac{1}{4}$	&0				&1 \\
			&$\Sigma K \eta K$		&1				&0				&0				&$\frac{3}{2}$ \\
			&$\Xi \pi K\eta$		&0				&1				&-2				&$\frac{1}{2}$ \\
			&$\Xi \eta K\pi$		&0				&1				&2				&$\frac{1}{2}$ \\

\hline
\hline
\end{tabular}
\end{center}
\caption{Coefficients of the Box Integral for the S=-1, I=0 states.}
\label{tab:intboxcoeff2}
\end{footnotesize}
\end{table}

\begin{table}
\begin{footnotesize}
\renewcommand{\arraystretch}{1.2}
\begin{center}
\begin{tabular}{c|c|c|c|c|c}
\hline
\multicolumn{6}{c}{Coefficients for state $S=-1,~I=1$}\\
\hline
\hline
Channel	&Box						&$(D+F)^2$		&$(D-F)^2$		&$(D+F)(D-F)$	&$C_{V_1}C_{V_2}$ \\
		&$B P_1 P_2 P_3$ & a & b & c & \\
\hline
$\bar{K}^* N$	&$\Sigma  K \pi K$	&0				&1				&0				&1 \\
				&$\Lambda K \pi K$	&1				&$\frac{1}{4}$	&-1				&$\frac{2}{3}$ \\
				&$N \pi K \pi$		&1				&0				&0				&$\frac{1}{2}$ \\
				&$\Sigma K \eta K$	&0				&1				&0				&$\frac{3}{2}$ \\
\hline
$\rho \Lambda$	&$\Sigma \pi\pi\pi$	&1				&1				&2				&$\frac{2}{3}$ \\
				&$\Xi KKK$			&$\frac{1}{4}$	&1				&-1				&$\frac{2}{3}$	\\
				&$N KKK$			&1				&$\frac{1}{4}$	&-1				&$\frac{2}{3}$	\\
\hline
$\rho \Sigma$	&$\Sigma \pi\pi\pi$	&1				&1				&-2				&1 \\
				&$\Lambda \pi\pi\pi$&1				&1				&2				&$\frac{2}{3}$	\\
				&$\Xi KKK$			&1				&0				&0				&1	\\
				&$N KKK$			&0				&1				&0				&1	\\
\hline
$\omega \Sigma$ &$\Xi KKK$			&1				&0				&0				&$\frac{1}{2}$	\\
				&$N KKK$			&0				&1				&0				&$\frac{1}{2}$	\\

\hline
$K^*\Xi$	&$\Sigma K \pi K$		&1				&0				&0				&$\frac{1}{2}$ \\
			&$\Xi \pi K\pi$			&0				&1				&0				&$\frac{5}{4}$ \\
			&$\Lambda K \pi K$		&$\frac{1}{4}$	&1				&-1				&1 \\
			&$\Xi \eta K\eta$		&1				&$\frac{1}{4}$	&0				&1 \\
			&$\Lambda K \eta K$		&$\frac{1}{4}$	&1				&-1				&2 \\
			&$\Xi \pi K\eta$		&0				&-1				&2				&$\frac{1}{2}$ \\
			&$\Xi \eta K\pi$		&0				&-1				&-2				&$\frac{1}{2}$ \\
\hline
$\phi \Sigma$	&$\Xi KKK$			&1				&0				&0				&1	\\
				&$N KKK$			&0				&1				&0				&1	\\
\hline
\hline
\end{tabular}
\end{center}
\caption{Coefficients of the Box Integral for the S=-1, I=1 states.}
\label{tab:intboxcoeff3}
\end{footnotesize}
\end{table}

\begin{table}
\begin{footnotesize}
\renewcommand{\arraystretch}{1.2}
\begin{center}
\begin{tabular}{c|c|c|c|c|c}
\hline
\multicolumn{6}{c}{Coefficients for state $S=-2,~I=1/2$}\\
\hline
\hline
Channel	&Box						&$(D+F)^2$		&$(D-F)^2$		&$(D+F)(D-F)$	&$C_{V_1}C_{V_2}$ \\
		&$B P_1 P_2 P_3$ & a & b & c & \\
\hline
$\bar{K}^*\Lambda$&$\Xi  K \pi K$		&$\frac{1}{4}$	&1				&-1				&1 \\
				&$\Sigma \pi K \pi$		&1				&1				&2				&$\frac{1}{4}$ \\
				&$\Xi K \eta K$			&$\frac{1}{4}$	&1				&-1				&1 \\
				&$\Lambda \eta K \eta$	&1				&1				&2				&$\frac{1}{4}$ \\
\hline
$\bar{K}^*\Sigma$&$\Xi  K \pi K$		&1				&0				&0				&$\frac{11}{12}$ \\
				&$\Sigma \pi K \pi$		&1				&1				&-2				&$1$ \\
				&$\Lambda \pi K \pi$	&1				&1				&2				&$\frac{1}{12}$ \\
				&$\Xi K \eta K$			&1				&0				&0				&$\frac{1}{4}$ \\
				&$\Sigma \eta K \eta$	&1				&1				&2				&$\frac{1}{4}$ \\
				&$\Sigma \pi K \eta$	&1				&-1				&0				&$-\frac{1}{3}$ \\
				&$\Sigma \eta K \pi$	&1				&-1				&0				&$-\frac{1}{3}$ \\
\hline
$\rho \Xi$		&$\Xi \pi\pi\pi$	&0				&1				&0				&4 \\
				&$\Lambda KKK$		&$\frac{1}{4}$	&1				&-1				&1	\\
				&$\Sigma KKK$		&1				&0				&0				&$\frac{1}{12}$	\\
\hline
$\omega \Xi$ 	&$\Lambda KKK$		&$\frac{1}{4}$	&1				&-1				&$\frac{1}{3}$	\\
				&$\Sigma  KKK$		&1				&0				&0				&$\frac{1}{4}$	\\
\hline
$\phi \Sigma$	&$\Lambda KKK$		&$\frac{1}{4}$	&1				&-1				&$\frac{2}{3}$	\\
				&$\Sigma  KKK$		&1				&0				&0				&$\frac{1}{2}$	\\
\hline
\hline
\end{tabular}
\end{center}
\caption{Coefficients of the Box Integral for the S=-2, I=1/2 states.}
\label{tab:intboxcoeff4}
\end{footnotesize}
\end{table}

\end{document}